\documentclass[%
reprint,
nofootinbib,
 aps,
]{revtex4-2}

\usepackage{graphicx}
\usepackage{dcolumn}
\usepackage{bm}
\usepackage{xcolor}
\usepackage{ulem}
\usepackage{booktabs}
\usepackage{multirow}
\usepackage{amsmath,amssymb}
\usepackage{hyperref}
\newcommand{\vew}{v_{\rm EW}}

\newcommand{\beq}{\begin{equation}}
\newcommand{\eeq}{\end{equation}}
\newcommand{\eq}[1]{Eq.~(\ref{#1})}

\begin{document}

\preprint{APS/123-QED}

\title{Probing the Electroweak Phase Transition in the Flipped Two-Higgs-Doublet Model at the LHC}

\author{Jidong~Du}
 \email{dujd@mail2.sysu.edu.cn}
 \affiliation{%
 School of Physics and Astronomy, Sun Yat-sen University (Zhuhai Campus), Zhuhai 519082, China
}%

\author{Yun~Jiang}%
 \email{jiangyun5@sysu.edu.cn (corresponding author)}
\affiliation{%
School of Physics and Astronomy, Sun Yat-sen University (Zhuhai Campus), Zhuhai 519082, China
}%
\affiliation{%
Guangdong Provincial Key Laboratory of Quantum Precision Measurement and Sensing, Zhuhai 519082, China
}%
\author{Wei~Su}
 \email{suwei26@mail.sysu.edu.cn}
\affiliation{%
School of Science, Shenzhen Campus of Sun Yat-sen University, No. 66, Gongchang Road, \\ Guangming District, Shenzhen, 518107, P.R. China 
}%

\date{\today}

\begin{abstract}
We study the CP-conserving flipped (Type-Y) Two-Higgs-Doublet Model (2HDM) in the large-$\tan\beta$ regime ($\tan\beta>30$), focusing on its implications for electroweak phase transitions (EWPTs) and LHC phenomenology. Viable parameter regions supporting a strong first-order EWPT fall into two heavy-Higgs hierarchies: (A) $m_{H^\pm}\simeq m_H<m_A$ and (B) $m_H<m_{H^\pm} \simeq m_A$, both featuring a heaviest CP-odd Higgs $A$. Scenario~A typically proceeds via one-step transitions with lower nucleation temperatures, while Scenario~B allows one-step or two-step transitions, opening the decay $A\to H^\pm W^\mp$ and yielding richer collider signatures. In all cases, nucleation conditions are satisfied, avoiding false-vacuum trapping. 
We assess LHC prospects through bottom-associated production with multi-$b$ final states: $pp\to bbH\to 4b$ and $pp\to bbA\to bb W^\pm H^\mp\to 4b\ell\ell\nu\nu$. The $4b$ channel offers high-statistics discovery potential, reaching signal significances $z\gtrsim 25$ at the 13 TeV LHC with 300 fb$^{-1}$ and up to $z\gtrsim 100$ at the 14 TeV HL-LHC with 3 ab$^{-1}$. The cascade channel, while experimentally more challenging, directly probes the heavy Higgs spectrum and can discriminate between EWPT scenarios. Using optimized selections with a BDT-based multivariate analysis, significances of $z \simeq 6.8$ can be achieved in favorable regions of Scenario~B at the HL-LHC. These results indicate that the HL-LHC can realistically probe the BSM Higgs sector responsible for a strong first-order EWPT and provide insight into the underlying phase transition dynamics in the flipped 2HDM.

\end{abstract}

\maketitle


\section{Introduction}
\label{intro}

The discovery of the Higgs boson at the LHC in 2012~\cite{ATLAS:2012yve,CMS:2012qbp} completed the particle content of the Standard Model (SM). While the SM provides an excellent description of collider data, it does not address several fundamental questions. Among the most prominent is the origin of the observed matter–antimatter asymmetry of the Universe~\cite{WMAP:2012fli}, which may be dynamically generated through electroweak baryogenesis (EWBG)~\cite{Morrissey:2012db}.

Successful EWBG requires the Sakharov conditions~\cite{Sakharov:1967dj}: baryon number violation, C and CP violation, and a departure from thermal equilibrium. The latter can be realized if the electroweak phase transition (EWPT) is strongly first order, commonly characterized by a transition strength $\xi  \gtrsim 1$~\cite{Moore:1998swa}. 
A strong first-order EWPT would also generate stochastic backgrounds of gravitational waves (GWs)~\cite{Caprini:2019egz}, potentially detectable by future space-based interferometers such as LISA~\cite{LISA:2017pwj}, Taiji~\cite{Hu:2017mde} and TianQin~\cite{TianQin:2015yph}.

In the SM, however, the EWPT is not first order. Nonperturbative studies~\cite{Kajantie:1996mn,Kajantie:1996qd,Csikor:1998eu,Aoki:1999fi} show that for a Higgs boson mass above approximately $70~{\rm GeV}$ the transition becomes a crossover, excluding EWBG within the SM. This motivates the study of beyond-SM (BSM) scenarios that modify the Higgs sector.

Among various BSM extensions, the Two-Higgs-Doublet Model (2HDM)~\cite{Branco:2011iw} provides a minimal and well-studied extension of the scalar sector. By introducing a second Higgs doublet, the model alters the structure of the Higgs potential and allows for the possibility of a strong first-order EWPT. Depending on the Yukawa structure, 2HDMs are classified into four types, each leading to different mass matrices and fermion coupling patterns after electroweak symmetry breaking (EWSB). 
It has been shown that Type-I and Type-II 2HDMs can accommodate a strong first-order EWPT and potentially generate successful  EWBG~\cite{Cline:1996mga,Fromme:2006cm,Cline:2011mm,Dorsch:2014qja,Dorsch:2016nrg,Basler:2016obg,Haarr:2016qzq,Basler:2017uxn,Bernon:2017jgv}. The extended scalar spectrum consists of a heavy CP-even scalar $H$, a CP-odd scalar $A$, and a charged Higgs boson $H^\pm$, and it can lead to distinctive collider signatures~\cite{Bernon:2017jgv,Kling:2020hmi,Goncalves:2022wbp}.

In this work we focus on the flipped (Type-Y) 2HDM. The lighter CP-even Higgs boson $h$ is identified with the observed $125~{\rm GeV}$ Higgs state~\cite{ATLAS:2012yve,CMS:2012qbp}. Its couplings to SM gauge bosons are controlled by $\cos(\beta-\alpha)$, which is constrained by precision Higgs measurements to be close to zero, $\cos(\beta-\alpha) \lesssim 0.3$~\cite{Haller:2018nnx}. In the alignment limit, $\cos(\beta-\alpha)=0$, the couplings of $h$ coincide with their SM values, while the interactions of the additional Higgs bosons depend only on $\tan\beta$. 

The Yukawa structure in the flipped 2HDM exhibits distinctive phenomenology at large $\tan\beta$, where couplings of $H$ and $A$ to down-type quarks are enhanced, leading to sizable production rates at colliders.
This contrasts with Type-I and Type-X models, where large $\tan\beta$ suppresses production, and with Type-II models, where $\tan\beta \gtrsim 25$ is strongly constrained by electroweak precision data and and flavor physics measurements for charged Higgs below $\sim 1~{\rm TeV}$~\cite{Haller:2018nnx}.

Motivated by these features, we investigate the EWPT in the flipped 2HDM, focusing on the large-$\tan\beta$ regime. We examine whether a strong first-order EWPT can be realized while satisfying current theoretical and experimental constraints, and investigate the associated collider phenomenology. 

The remainder of this paper is organized as follows. Section~\ref{sec:Model_des} provides a description of the flipped 2HDM. Section~\ref{sec:scan} describes the parameter scan and the resulting viable points. In Section~\ref{sec:ewpt}, we analyze the EWPT, while Section~\ref{sec:Collider} is devoted to collider phenomenology. Our conclusions are presented in Section~\ref{sec:conc}.

\section{Model Framework}
\label{sec:Model_des}

We begin with a brief review of the CP-conserving flipped (Type-Y) 2HDM at tree-level and zero temperature. The model contains two scalar $SU(2)_L$ doublets, $\Phi_i,i=1,2$, with the same quantum numbers as the SM Higgs field. 

To prevent tree-level Higgs-mediated flavor-changing neutral currents (FCNCs), a discrete $\mathbb{Z}_2$ symmetry is imposed under which $\Phi_2 \to -\Phi_2$ while $\Phi_1$ remains invariant. 
Allowing for a soft breaking of this symmetry, the most general renormalizable and CP-conserving scalar potential is given by
\beq
\begin{aligned}
	\label{eq:V0}
	V_{\rm tr}&=m_{11}^2\Phi_1^\dagger\Phi_1+m_{22}^2\Phi_2^\dagger\Phi_2-\left( m_{12}^2\Phi_1^\dagger\Phi_2+h.c.\right) \\
	&+\frac{1}{2}\lambda_1\big(\Phi_1^\dagger\Phi_1\big)^2+\frac{1}{2}\lambda_2\big(\Phi_2^\dagger\Phi_2\big)^2 \\
	&+\lambda_3\big(\Phi_1^\dagger\Phi_1\big)\big(\Phi_2^\dagger\Phi_2\big)+\lambda_4\big(\Phi_1^\dagger\Phi_2\big)\big(\Phi_2^\dagger\Phi_1\big)\\
	&+ \frac{1}{2} \left[ \lambda_5\big(\Phi_1^\dagger\Phi_2\big)^2+h.c. \right] 
\end{aligned}
\eeq
where all parameters in the potential are taken to be real.

Electroweak symmetry breaking (EWSB) occurs when the scalar mass matrix develops at least one negative eigenvalue, inducing nonzero vacuum expectation values (vevs) for both Higgs doublets.
We parametrize the doublets as,
	\begin{equation}
		\label{Phi_def}
		\Phi_i=\frac{1}{\sqrt{2}}\begin{pmatrix}
			\phi_{i}^{\pm}\\
			(v_i+\rho_i+i\eta_i)
		\end{pmatrix}, \,\, i=1,2
	\end{equation}
where $v_i$ denote the vevs of the two doublets, $\rho_i$ are the CP-even components, and $\eta_i$ are the CP-odd components. Any complex phase in the vevs can be rotated away without loss of generality, so both $v_i$ are taken to be real.
		
The scalar potential, \eq{eq:V0}, contains eight parameters: $ m_{11}^2, m_{22}^2, m_{12}^2, \lambda_{1,2,3,4,5} $. 
The quadratic mass parameters $m^2_{11}$ and $m^2_{22}$ can be eliminated using the minimization conditions,
	\begin{flalign}
		m_{11}^2&=m_{12}^2\tan\beta-\frac{1}{2} v ^2(\lambda_1 \cos^2\beta+\lambda_{345}\sin^2\beta)\label{m11_def} \\
		m_{22}^2&=m_{12}^2\cot\beta+\frac{1}{2} v^2(\lambda_2 \sin^2\beta+\lambda_{345}\cos^2\beta)\label{m22_def}
	\end{flalign}
with $\lambda_{345}\equiv \lambda_3+\lambda_4+\lambda_5$. The vevs are expressed in terms of
\beq
v^2\equiv v_1^2+v_2^2, \quad \tan\beta=v_2/v_1.
\eeq 
where $v=v_{\rm EW}=246~{\rm GeV}$, corresponding to the zero-temperature electroweak vev. 
After applying the minimization conditions, the scalar potential is specified by seven independent parameters, which may be chosen as: $ \{\tan\beta, m_{12}^2, \lambda_{1,2,3,4,5}\} $.
	
Among the eight scalar degrees of freedom, three Goldstone bosons, $G^\pm= \cos\beta \phi^\pm_1 + \sin\beta \phi^\pm_2$ and $G^0 = \cos\beta \eta_1 + \sin\beta \eta_2$, are absorbed by the $W^\pm$ and $Z$ bosons via EWSB. 
The remaining five physical states consist of a CP-odd scalar $A$, a pair of charged Higgs bosons $H^\pm$ and two CP-even Higgs scalars $h$ and $H$ with $m_h<m_H$. These states are defined by 
	\begin{align}
		\label{Phy_def}
		A &= -\sin\beta \eta_1 + \cos\beta \eta_2, \\
		H^\pm &= -\sin\beta \phi^\pm_1 + \cos\beta \phi^\pm_2,\\
		\begin{pmatrix}
			h\\H
		\end{pmatrix}&=
		\begin{pmatrix}
			-\sin\alpha&\cos\alpha\\
			\cos\alpha&\sin\alpha
		\end{pmatrix} 
		\begin{pmatrix}
			\rho_1\\\rho_2
		\end{pmatrix}
	\end{align}
where $\alpha$ is mixing angle that diagonalizes the mass matrix of the CP-even states. The masses of the CP-odd and charged Higgs bosons are given by
	\begin{align}
	m^2_A &= -\lambda_5 \vew^2+2m^2_{12}/\sin(2\beta) \label{ma_def}, \\
	m^2_{H^\pm} & = m^2_A - {1\over 2} \vew^2(\lambda_4-\lambda_5)\label{mhc_def}.
	\end{align}
while the explicit expressions for $m_h$ and $m_H$ can be found in Ref.~\cite{Branco:2011iw}.

The Yukawa sector depends on how fermions couple to the two Higgs doublets. In the flipped (Type-Y) model considered here, up-type quarks and charged leptons couple to $\Phi_2$, while down-type quarks couple to $\Phi_1$. This structure ensures the absence of tree-level FCNCs. 
The Yukawa Lagrangian in the flipped 2HDM is
	\beq
	\begin{aligned}\label{L_yuk}
		\mathcal{L}_{\rm Yuk} =-\sum_{i, j} & \Big(\bar{Q}_{i L} \widetilde{\Phi}_2 Y_{i j}^u u_{j R}+\bar{Q}_{i L} \Phi_1 Y_{i j}^d d_{j R}\\
		& \quad +\bar{L}_{i L} \widetilde{\Phi}_2 Y_{i j}^l \ell_{j R}\Big)+\text { h.c. }
	\end{aligned}
	\eeq
	
After EWSB, it can be written in terms of the physical Higgs fields as:
	\beq
	\begin{aligned}
		\mathcal{L}_{\rm Yuk}= & -\sum_{f=u, d, \ell} y_f^{\rm SM} \left(\xi_f^h \bar{f} f h+\xi_f^H \bar{f} f H + i \zeta_f \xi_f^A \bar{f} \gamma_5 f A\right) \\
		& -\Big\{ \bar{u} V_{u d} \left(y^{\rm SM}_u \xi_u^A P_L+y^{\rm SM}_d \xi_d^A P_R\right) d H^{+}\\
		&\quad\quad +y_l^{\rm SM} \xi^A_l \bar{\nu}_L l_R H^{+}+\text {h.c.}\Big\} 
	\end{aligned}
	\eeq
where $y_f^{\rm SM}=\sqrt{2}m_f/\vew$ denotes the SM Yukawa coupling for fermion $f=u,d,l$. For the CP-odd Higgs couplings, $\zeta_d=-1$ and $\zeta_{u,l}=+1$. The explicit expressions of $\xi_f^{h,H,A}$ are summarized in Table~\ref{Yukawa_tab}. These factors encode the dependence of the Higgs–fermion interactions on the mixing angles $\alpha$ and $\beta$, and determine the characteristic $\tan\beta$ scaling behavior in the alignment limit, which will be relevant for our analysis.

For phenomenological studies, it is convenient to trade the quartic couplings $\lambda_{1,2,3,4,5}$ for observable parameters. Using the relation between the scalar potential parameter and physical Higgs masses $m_{h,H,A,H^\pm}$ and the mixing angle $\alpha$~\cite{Dubinin:1998nt}, the model can be fully specified by the independent set 
$$\{ m_h, m_H, m_A, m_{H^\pm}, \tan\beta, \cos(\beta-\alpha), m_{12}^2\}.$$ 
This complete parameter basis will be used in the numerical analysis presented in the next section.

	\begin{table}[t]
		\caption{Normalized Yukawa couplings of the neutral Higgs bosons ($h$, $H$, $A$) to up-type quarks, down-type quarks, and charged leptons in the Type-Y (flipped) 2HDM~\cite{Branco:2011iw}.}
		\label{Yukawa_tab}
		\centering
		\begin{tabular}{cccc}
			\hline  $f=$&$u$&$d$&$l$\\
			\hline  $\xi^h_f$ & $\cos\alpha/\sin\beta$ & $-\sin\alpha/\cos\beta$ & $\cos\alpha/\sin\beta$ \\
			\hline  $\xi^H_f$ & $\sin\alpha/\sin\beta$ & $\cos\alpha/\cos\beta$ & $\sin\alpha/\sin\beta$ \\
			\hline  $\xi^A_f$ &$\cot\beta$ & $-\tan\beta$ & $-\cot\beta$ \\
			\hline
		\end{tabular}
	\end{table}

\section{Parameter Scan and Constraints}
\label{sec:scan}
	
We identify the lighter CP-even Higgs boson $h$ as the SM-like Higgs boson observed at the LHC and fix its mass to $m_h=125.09~{\rm GeV}$~\cite{ATLAS:2012yve,CMS:2012qbp}.
Since $\cos(\beta-\alpha)$ governs the coupling of $h$ to the electroweak gauge bosons and quantifies deviations from the SM limit, it is strongly constrained by Higgs precision measurements. 

Guided by global analyses in Type-II models~\cite{Bernon:2015qea,Bernon:2015wef,Dorsch:2016tab,Karmakar:2019vnq}, we impose $\cos(\beta-\alpha) \lesssim 0.09$ ensuring near alignment.  

In this regime, the effective couplings of the heavy Higgs bosons $H$, $A$ and $H^{\pm}$ to fermions, $\xi_f$ (see Table~\ref{Yukawa_tab}), depend predominantly on $\tan\beta$. To be specific, the couplings of $H$ and $A$ to down-type quarks are enhanced at large $\tan\beta$, leading to increased production cross sections and enhanced branching ratios into $b\bar b$.

Motivated by this feature, we focus on $30 \leq \tan\beta \leq 42$. 
Large $\tan\beta$ values are, however, strongly constrained by flavor observables and direct LHC searches for BSM Higgs bosons. In particular, the radiative decay $B \to X_s \gamma$ imposes a lower bound on the charged Higgs mass in Type-II/Y 2HDM~\cite{Li:2024kpd,Atkinson:2021eox,Misiak:2020vlo,Misiak:2017bgg,Arbey:2017gmh}. 
While a stronger bound $m_{H^\pm}<800~{\rm GeV}$ has been quoted\footnote{We note that imposing the stronger bound $m_{H^\pm} \gtrsim 800~\mathrm{GeV}$ generally precludes a simultaneous explanation of other flavor anomalies within the Type-II/Y 2HDM~\cite{Atkinson:2021eox,BaBar:2012obs}.}, we adopt the commonly used conservative limit $m_{H^\pm} \gtrsim 580\,\mathrm{GeV}$ following Refs.~\cite{Haller:2018nnx,Misiak:2017bgg}. 

The heavy scalar masses $m_H$ and $m_A$ are treated as free scan parameters within the ranges specified in Table~\ref{scan_range_tab}. For each parameter point, the soft-breaking parameter $m_{12}^2$ is determined dynamically such that vacuum stability and perturbativity are maintained once the remaining parameters are fixed.

	\begin{table}[t]
		\caption{Ranges of the scalar masses, mixing angles, and $\tan\beta$ used in the scan.\label{scan_range_tab}}
		\centering
		\begin{tabular}{ccccc}
			\hline  $\tan\beta$&$\cos(\beta-\alpha)$&$m_H$&$m_A$&$m_{H^\pm}$ \\
			\hline  $[30,42]$&$[0,0.09]$&$[130,1000]$&$[62.5,1000]$&$[580,1000]$ \\
			\hline
		\end{tabular}
	\end{table} 
 
We perform a random scan using \texttt{2HDMC}~\cite{Eriksson:2009ws} and \texttt{HiggsTools}~\cite{Bahl:2022igd}, imposing both theoretical consistency conditions and experimental constraints. 
The theoretical consistency conditions include: 
	
\textbf{Vacuum stability}: The scalar potential must be bounded from below to ensure the existence of a stable electroweak vacuum. This requirement is satisfied if the potential remains positive in all directions of field space at asymptotically large field values. 
The necessary and sufficient conditions at tree level are
\beq
\label{eq:stab}
\lambda_1,  \lambda_2> 0, \,\, \lambda_3 > -\sqrt{\lambda_1 \lambda_2}, \,\, \lambda_3 + \lambda_4 - |\lambda_5| > -\sqrt{\lambda_1 \lambda_2} 
\eeq
	
\textbf{Perturbative unitarity}: The $S$-matrix must satisfy unitarity for all $2\to2$ scalar scattering processes, including channels involving Higgs and Goldstone bosons. The partial-wave amplitudes $a_j$ for angular momentum $j$ are required to satisfy
\beq
\label{eq:unit}
|a_j| \leq 1, \,\, 0\leq {\rm Im}(a_j) \leq 1, \,\, |{\rm Re}(a_j)|\leq {1\over 2}
\eeq 
	
\textbf{Tree-level perturbativity}: To ensure reliable perturbative predictions, we further require the quartic couplings to satisfy $|\lambda_i| < 4\pi$, 
This prevents excessively large tree-level amplitudes that could otherwise violate Eq.~(\ref{eq:unit}).

In fact, these last two conditions constrain combinations of the quartic couplings and prevent the breakdown of perturbative unitarity at tree level.
	
The constraints from experimental measurement at low-energy include:
	
\textbf{Electroweak precision observables (oblique parameters)}: 
New scalar states contribute to the vacuum polarizations of electroweak gauge bosons, parametrized by the oblique parameters $S$, $T$, and $U$. We adopt the global-fit results from Ref.~\cite{PhysRevD.110.030001} with correlation coefficients from Ref.~\cite{Haller:2018nnx}:
    \beq
    S = -0.04 \pm 0.1, \,\,  T = 0.01 \pm 0.12, \,\, U = -0.01 \pm 0.09.
    \eeq
For each parameter point we compute $\chi^2_{\rm STU}$ and require $\chi^2_{\rm STU} < 5.99$, corresponding to the 95\% confidence level~(CL). 
	
\textbf{Precision Higgs measurements}: 
Compatibility with Higgs data is evaluated using the \texttt{HiggsSignals} module~\cite{Bechtle:2020uwn} within \texttt{HiggsTools}. The global fit includes 131 observables, comprising the Higgs mass, signal strengths, and coupling measurements. For each parameter point, the fit yields a $\chi^2_{\rm Higgs}$ value. We impose
$\chi_{\rm Higgs}^2 < 158.7$, which corresponds to the 95\%~CL for 131 degrees of freedom.
	
\textbf{Direct searches for BSM Higgs bosons:} Constraints from LHC searches for BSM Higgs states are implemented using the \texttt{HiggsBounds} module~\cite{Bechtle:2020pkv}. For each parameter point, we compute 
\beq
\mu_\mathrm{obs}=\frac{(\sigma\cdot\mathrm{BR})_\mathrm{the}}{(\sigma\cdot\mathrm{BR})_\mathrm{95\%~CL. exp}}.
\eeq
If any search channel yields $\mu_\mathrm{obs}>1$, the parameter point is excluded at 95\% CL. 

The implementation covers a wide range of LHC search channels. In our analysis, the most constraining channels typically include:
	\begin{itemize}
		\item $\rm bbH$ with $H \rightarrow b\bar{b}$~\cite{ATLAS:2019tpq,CMS:2018hir}
		\item $pp \to H/A \rightarrow AZ/HZ \rightarrow b\bar{b}\ell^+\ell^-$~\cite{ATLAS:2018oht,CMS:2019ogx}
		\item $pp \to A\to Zh\to \ell^+\ell^-\tau^+\tau^-$~\cite{CMS:2019kca}
		\item $\rm bbA$ with $A\to Zh\to Zb\bar{b}$~\cite{CMS:2019qcx}
		\item $pp \to A/H\to ZH/A\to \ell^+\ell^-W^+W^-/\ell^+\ell^-b\bar{b}$~\cite{ATLAS:2020gxx}
        \item $\rm bbH$ with $H \to hh \to bb\bar{b}\bar{b}$~\cite{ATLAS:2018rnh}
	\end{itemize}
These direct searches provide some of the strongest constraints on the heavy scalar mass spectrum, particularly in regions with enhanced bottom-Yukawa couplings at large $\tan\beta$.

\begin{figure}[t]
\centering
\includegraphics[width=0.45\textwidth]{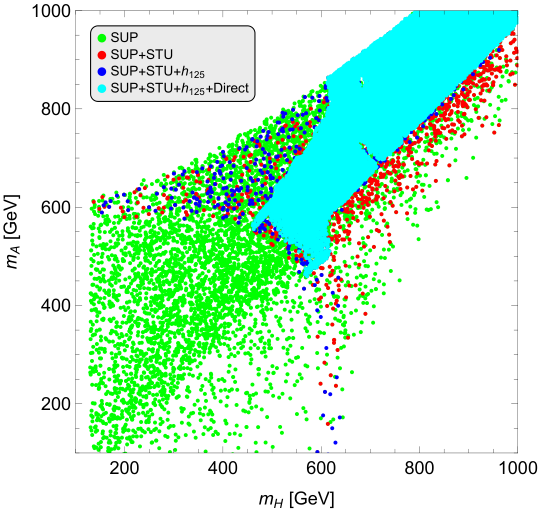}
\caption{Points generated in the parameter scan shown in the $m_H$ vs.\ $m_A$ plane. The color indicate the sequential imposition of constraints: stability + unitarity + perturbativity (SUP) in green, SUP + electroweak precision observables (STU) in blue, SUP + STU + Higgs precision data ($h_{125}$) in red, and SUP + STU + $h_{125}$ + direct searches for BSM Higgs bosons (Direct) in cyan. Each layer represents the points surviving up to that stage, with the cyan points corresponding to the fully allowed parameter space after all constraints are applied. The density of points differs between layers because only a representative subset of excluded points is retained at each step of the scan.
\label{fig:scan}}
\end{figure}
	
The results of our parameter scans are displayed in Fig.~\ref{fig:scan}. 
Theoretical constraints, particularly perturbativity, significantly restrict large mass splittings between $H$ and $A$, excluding regions with sizable $|m_H - m_A|$. 
Electroweak precision data, mainly through the $T$ parameter, further disfavor spectra in which both $m_H$ and $m_A$ lie below about $500~{\rm GeV}$. 

Higgs precision measurements, implemented through \texttt{HiggsSignals}, primarily constrain $\cos(\beta-\alpha)$ and indirectly restrict correlated regions of the heavy scalar spectrum. Although near alignment is imposed explicitly in our scan, we observe that regions with large negative mass splittings $\Delta m \equiv m_A - m_H \lesssim -50~\mathrm{GeV}$ are disfavored once all constraints are applied. 
Importantly, this boundary does not arise from a direct kinematic threshold in Higgs decays. Instead, it emerges from the global fit to the 125~GeV Higgs signal strengths.
When theoretical constraints on the quartic couplings and electroweak precision data are simultaneously imposed, the allowed mass relations among the heavy scalars become highly restricted. Within this constrained parameter space, Higgs precision measurements effectively forces $\cos(\beta - \alpha)$ to lie extremely close to zero. In this situation, variations in the heavy Higgs spectrum---particularly sizable mass splittings between $H$ and $A$---induce correlated shifts in the Higgs couplings through the underlying relations of the scalar potential. Consequently, even if $\cos(\beta - \alpha)$ is treated as an independent scan parameter, large negative $\Delta m$ values tend to drive the model away from the level of alignment required by the global Higgs fit, leading to their exclusion. 

An exception appears as a narrow allowed strip around $m_H \simeq 600~\mathrm{GeV}$, corresponding to the so-called ''wrong-sign'' Yukawa regime. In this region, the 125~GeV Higgs coupling to down-type quarks has the opposite sign relative to the SM prediction while retaining nearly SM-like magnitudes. Loop-induced processes such as $h\to\gamma\gamma$ and gluon fusion allow partial cancellations between fermion and gauge contributions, so percent-level deviations from exact alignment remain compatible with current Higgs precision measurement data.

Direct searches for additional Higgs bosons, implemented via \texttt{HiggsBounds}, impose the strongest constraints in the $(m_H, m_A)$ plane. 
Channels involving $\rm bbA/H$ production and cascade decays such as $A/H\to ZH/A, b\bar b$~\cite{ATLAS:2018oht,CMS:2019ogx,CMS:2019qcx,ATLAS:2019tpq,CMS:2018hir} exclude large regions characterized approximately by $m_A - 2m_H \gtrsim 350~\mathrm{GeV}$, generating an effective lower bound on $m_A$ that increases roughly linearly with $m_H$. These searches also carve out a small excluded pocket near $(m_H, m_A) \simeq (700, 850)\mathrm{GeV}$. 
Additional exclusions appear around $(m_H,m_A)\simeq(450-550,450-550),\mathrm{GeV}$ from multi-bottom final states~\cite{ATLAS:2018rnh} and near $(700,700)\mathrm{GeV}$ from $\rm bbH/A$ searches~\cite{ATLAS:2019tpq,CMS:2018hir}.

The sharp and irregular exclusion boundaries reflect the discrete mass grids used in experimental analyses. Overall, the surviving parameter points correspond to comparatively heavy BSM Higgs spectra with controlled mass splittings and near alignment. These viable regions provide the physically consistent starting point for the finite-temperature analysis presented in the next section.
	
\section{Electroweak Phase Transition}
\label{sec:ewpt}
	
\subsection{One-Loop Finite-Temperature Effective Potential}
\label{sec:Veff}

To investigate the EWPT, we construct the one-loop effective potential at finite temperature,
\beq
	V_{\rm eff}(\{\phi\})=V_0+V_1+V_{\rm CT}+V_T+V_{\rm daisy}\label{V_eff}
\eeq
where $\{\phi\}$ denotes the background classical fields. 
Restricting to CP-conserving field configurations, we take $\{\phi\}=\{\rho_1,\rho_2\}$. 
The tree-level potential $V_0$ is given by 
\beq
\begin{aligned}
\label{V0}
	V_{0}(\rho_1,\rho_2)&=\frac{1}{2}m_{11}^2\rho_1^2+\frac{1}{2}m_{22}^2\rho_2^2-m_{12}^2\rho_1 \rho_2 \\
			&+\frac{1}{8}\lambda_1\rho_1^4+\frac{1}{8}\lambda_2\rho_2^4+\frac{1}{4}(\lambda_3+\lambda_4+\lambda_5)\rho_1^2\rho_2^2 
\end{aligned}
\eeq

$V_1$ denotes the zero-temperature one-loop correction, which in the ($\overline{\text{MS}}$) scheme reads
\beq
	V_1(\rho_1,\rho_2) = \sum_i \frac{(-1)^{2s_i} n_i}{64\pi^2} m_i^4(\rho_1,\rho_2) \left[ \ln \frac{m_i^2(\rho_1,\rho_2)}{\mu^2} - c_i \right] \label{V1}
\eeq
where the sum runs over all particle species with field-dependent squared masses $m_i^2(\rho_1,\rho_2)$. Here $s_i$ denotes the spin and $n_i$ is the number of internal degrees of freedom: 
\begin{itemize}
\item Neutral scalars $\{h,H,A,G^0\}$: $n_i=1$;
\item Charged scalars $\{H^\pm, G^\pm\}$: $n_i=2$;
\item Gauge bosons: $n_i = 1$ for $Z_L$ (the longitudinal $Z$ boson); $n_i = 2$ for $W^\pm_L$ (longitudinal $W^\pm$) and for $Z_T$ (transverse $Z$); $n_i = 4$ for $W^\pm_T$ (transverse $W^\pm$);
\item Fermions: $n_i = 4$ for leptons and $n_i = 12$ for quarks.
\end{itemize}
The renormalization scale is denoted by $\mu$. In the $\overline{\rm MS}$ scheme, $c_i = 1/2$ for transverse gauge bosons and $c_i = 3/2$ for scalars, fermions, and longitudinal gauge bosons. 

Including $V_1$ shifts the vacuum away from its tree-level location and violates the tree-level minimization conditions, Eqs.~(\ref{m11_def}, \ref{m22_def}). To preserve the Higgs vevs $v_1, v_2$ and the physical scalar masses at zero temperature, we introduce counterterms,
\beq
\begin{aligned}
	\label{VCT}
		V_{\rm CT}(\rho_1, \rho_2) &= \delta m^2_{ 11} \rho^2_1 + \delta m^2_{22} \rho^2_2 + \delta m^2_{12} \rho_1 \rho_2 \\
		&+ \delta \lambda_1 \rho^4_1 + \delta \lambda_2 \rho^4_2 + \delta \lambda_{345} \rho^2_1 \rho^2_2.
\end{aligned}
\eeq
following the prescription of Ref.~\cite{Bernon:2017jgv,Cline:2011mm}.		
			
The finite-temperature contribution is given by~\cite{Arnold:1992rz}
\beq
V_T(\rho_1,\rho_2,T) = \sum_{i} n_i  \frac{T^4}{2\pi^2}  J_{\pm}\left(\frac{m_i^2(\rho_1,\rho_2)}{T^2}\right),
\label{V_T}
\eeq
where
\beq
J_{\pm}(y^2) = \mp \int_0^{\infty} \! dx  \,x^2 \ln \left[ 1 \pm e^{-\sqrt{x^2 + y^2}} \right],
\label{spe_J}
\eeq
with the upper (lower) sign corresponding to fermions (bosons). 

To account for infrared-sensitive bosonic modes, we include the daisy (ring) resummation~\cite{Arnold:1992rz,Carrington:1991hz},
\beq
\label{V_daisy}
\begin{aligned}
	V_{\rm daisy}(\rho_1,\rho_2,T)=&-\frac{T}{12\pi}\sum_i n_i[(M_i^2(\rho_1,\rho_2,T))^{3/2}\\
	&\quad\quad\quad\quad\quad\quad -(m_i^2(\rho_1,\rho_2))^{3/2}]
\end{aligned}
\eeq
where the sum includes all bosonic degrees of freedom except transverse gauge modes. At finite temperature, they acquire thermal (Debye) masses.$M_i^2(\rho_1,\rho_2,T)= m_i^2(\rho_1,\rho_2)+\Pi_i(T)$
with $\Pi_i(T)$ denoting the Debye self-energies.

Two commonly used resummation prescriptions are the Parwani~\cite{Parwani:1991gq} and Arnold–Espinosa schemes~\cite{Arnold:1992rz}. In the Parwani approach, all bosonic masses in $V_1$ and $V_T$ are replaced by their thermally corrected masses. 
We instead adopt the Arnold–Espinosa prescription, in which Debye corrections enter only through the cubic term via $V_{\rm daisy}$. 
This avoids overcounting of higher-order thermal effects and is generally considered more reliable for EWPT studies~\cite{Basler:2016obg,Cline:1996mga,Cline:2011mm}.
			
With the effective potential specified, we analyze the thermal evolution of the vacuum structure using \texttt{PhaseTracer}\cite{Athron:2020sbe}. 
At high temperature, the Universe resides in the symmetric phase with $\rho_1=\rho_2=0$. 
As it cools, a second minimum with $(\rho_1,\rho_2)\neq(0,0)$ develops, corresponding to the electroweak–broken phase, and eventually becomes energetically favored. 
For a first-order EWPT, the transition from the symmetric phase to the broken one proceeds via thermal tunneling through bubble nucleation. The nucleation rate per unit volume is approximately
\beq
\Gamma(T) \approx T^4 \left(\frac{S_3}{2\pi T}\right)^{3/2} e^{-\frac{S_3}{T}}\label{Tunl_rate}
\eeq
where $S_3$ is the three-dimensional Euclidean action evaluated on the bounce solution. 
At the critical temperature $T_c$, the symmetric and broken phases are degenerate in free energy and $S_3$ diverges, suppressing nucleation. As the temperature decreases below $T_c$, $S_3/T$ decreases and nucleation becomes efficient.
The phase transition effectively occurs at the nucleation temperature $T_n$, defined by the condition that roughly one bubble nucleates per Hubble volume. 
For EWPTs, this typically corresponds to
\beq
    S_3(T_n)/T_n \approx 140.
\eeq

\subsection{Results}
\label{sec:ewpt_res}

\begin{figure}[t]
\centering
\includegraphics[width=0.95\linewidth]{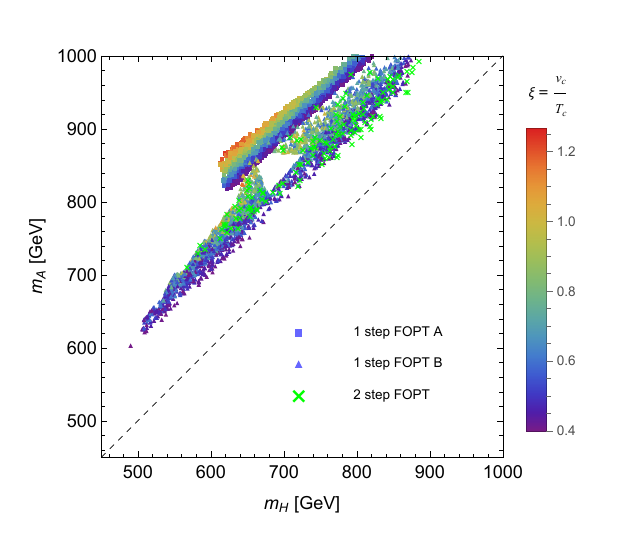}
\caption{EWPT results in the $m_H$ vs.\ $m_A$ plane, where the color denotes the phase transition strength $\xi \equiv v_c / T_c$. Only points with $\xi > 0.4$ are displayed to ensure successful EWPT via bubble nucleation. 
All viable points feature the typical hierarchy $m_A > m_H$ and satisfy the nucleation upper bound $\xi < 1.8$. Larger $\xi$ values  correlate with sizable mass splittings between $H$ and $A$, reflecting enhanced bosonic thermal contributions. Two distinct clusters with different mass hierarchies are clearly visible, corresponding to Scenario A (boxes) and Scenario B (triangles) as defined in the text, while two-step transitions (marked by crosses) only occurs in Scenario B.
\label{PT_Res}}
\end{figure}

The properties of the successful EWPT points are shown in Fig.~\ref{PT_Res}, where the color encodes the transition strength $\xi \equiv v_c/T_c$. We require $\xi > 0.4$, below which thermal tunneling is too inefficient for successful completion of the transition. 
A striking feature of the viable first-order EWPT region is a sizable mass splitting between the two neutral BSM Higgs bosons, $\Delta m \gtrsim m_Z$. This behavior follows directly from the structure of the finite-temperature effective potential. Bosonic thermal corrections generate cubic terms scaling as $m_\Phi^3(T)$, which are responsible for generating the barrier separating the symmetric and broken phases. 
In this model, these contributions arise primarily from the heavy scalars $H$, $A$, and $H^\pm$, whose effects are governed by their masses---or equivalently by the scalar couplings $\lambda_{3,4,5}$ that determine the mass splittings~\cite{Curtin:2016urg,Bernon:2017jgv}. 
When $\Delta m\simeq 0$, the heavy scalar contributions partially cancel, suppressing the barrier and preventing a first order EWPT. Once $\Delta m$ exceeds $\mathcal{O}(m_Z)$---the same scale that governs the SM gauge-induced cubic term---the additional scalar contributions sufficiently enhance the barrier to trigger a first-order EWPT, and accordingly the phase transition strength $\xi$ increases with $\Delta m$, as seen in Fig.~\ref{PT_Res}.

Two distinct clusters of viable points are visible in the figure, their origin will be clarified shortly. 
One-step phase transitions occur in both clusters of the allowed parameter space, whereas two-step transitions arise only in one of them. In the following, we focus on the one-step case for clarity. All viable one-step FOPT points satisfy $\xi < 1.3$, well below the nucleation upper bound $\xi<1.8$ derived in Ref.~\cite{Biekotter:2022kgf} for the Type-II 2HDM, and we have explicitly verified that all point with $\xi > 0.4$ successfully complete bubble nucleation. 
A triangular region around $m_H \sim 700~\text{GeV}$ and $m_A \sim 850~\text{GeV}$ is excluded by searches for $pp \to H/A \to AZ/HZ $, where the $Z$ boson decays leptonically and the accompanying Higgs decays to a bottom-quark pair~\cite{ATLAS:2018oht, CMS:2019ogx}.\footnote{The exclusion is less visible in Fig.~\ref{fig:scan} due to overlap between the two viable clusters, but becomes evident in Fig.~\ref{PT_Res} once the condition $\xi>0.4$ is imposed.}  
			
\begin{figure}[t]
\centering
\label{res_xigtpt8.a}\includegraphics[width=0.95\linewidth]{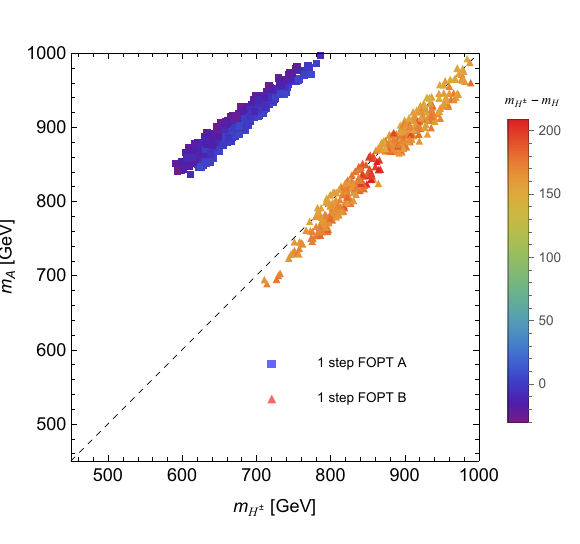}\\
\label{res_xigtpt8.b}\includegraphics[width=0.95\linewidth]{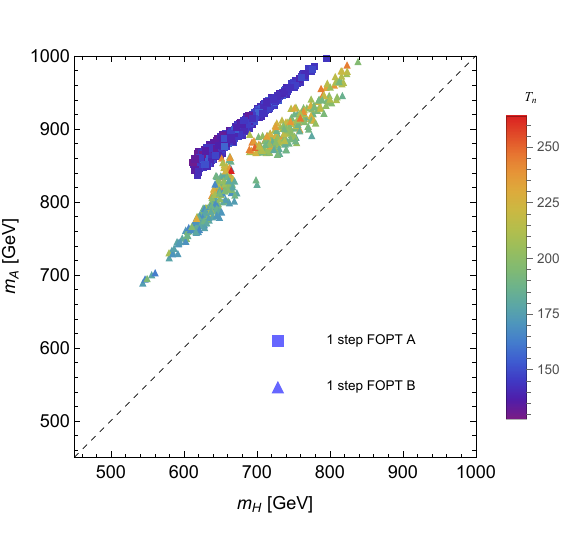}
\caption{(\textbf{Top}) Mass relations for strong first-order EWPT points with $\xi > 0.8$ shown in the $m_{H^\pm}$ vs.\ $m_A$ plane. A quasi-linear correlation between $m_{H^\pm}$ and $m_A$ emerges from the requirement that the charged Higgs be nearly degenerate with either $H$ or $A$, leading to two distinct mass hierarchies corresponding to Scenario A (boxes) and Scenario B (triangles). 
(\textbf{Bottom}) The corresponding nucleation temperature $T_n$ for the same $\xi > 0.8$ points, separated into Scenario A and Scenario B, showing a clear systematic difference between the two cases due to their distinct thermal dynamics and barrier structures.
\label{res_xigtpt4}}
\end{figure}

To identify scenarios relevant for EWBG and potentially observable GW signals, we further impose $\xi > 0.8$. Achieving such a strong transition requires a sizable thermal cubic term, dominantly generated by the heavy BSM Higgs bosons. This, in turn, implies a substantial mass splitting, $\Delta m \gtrsim \mathcal{O}(10^2)\,\mathrm{GeV}$, as shown in Fig.~\ref{res_xigtpt4}. 
Within this region, two characteristic mass hierarchies among the BSM Higgs bosons emerge:
\begin{itemize}
\item Scenario A: $m_{H^\pm} \simeq m_H$, with $A$ the heaviest state;
\item Scenario B: $m_{H^\pm} \simeq m_A$, with $H$ the lighter neutral state.
\end{itemize}

A more quantitative inspection shows that realizing $\xi > 0.8$ typically requires $\Delta m \gtrsim 150~(200)~\mathrm{GeV}$ in Scenario A (B). This threshold follows from the scaling of the bosonic cubic contribution, $V^{\rm cubic}_T \sim -T\, m^3(T)$. After including Debye corrections from $V_{\rm daisy}$, the enhancement becomes significant once $\Delta m$ reaches $\mathcal{O}(10^2)\,\mathrm{GeV}$, consistent with the lower boundary of the strong first-order EWPT region identified in our scan.

Although both scenarios allow $\xi>0.8$, their thermal histories differ. As shown in Fig.~\ref{res_xigtpt4}, the nucleation temperature $T_n$ is systematically lower in Scenario~A than in Scenario~B. This reflects how the different BSM scalar mass spectra shape the temperature-dependent potential barrier, leading to nucleation at distinct temperatures. 
	
\section{LHC Probes of a Strong EWPT: Monte Carlo Analysis}
\label{sec:Collider}

\begin{figure}[t]
\centering
\includegraphics[width=0.8\linewidth]{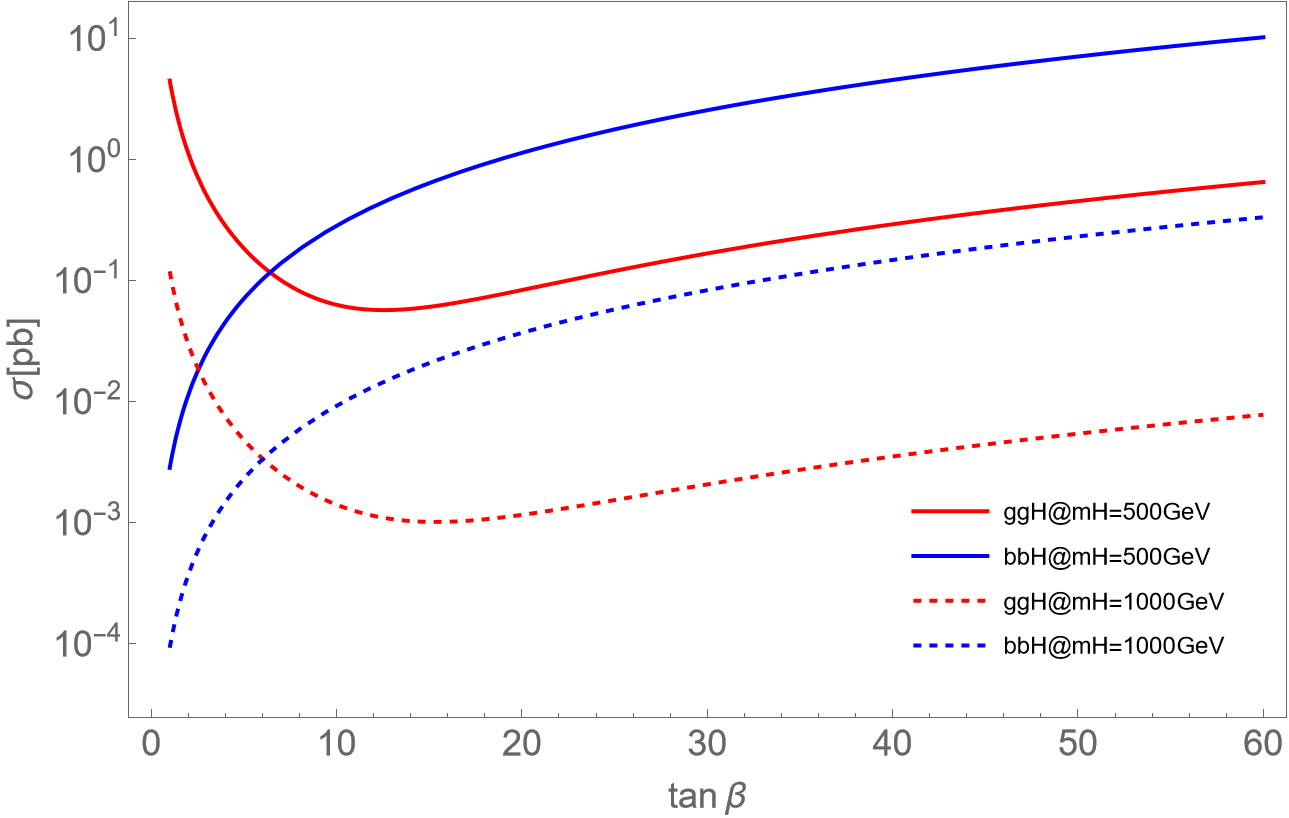}
\caption{Production cross sections at which $\sqrt{s}=13~{\rm TeV}$ for gluon fusion (red) and bottom-associated production $\rm bbH$ (blue) as functions of $\tan\beta$ for $m_H=500$~GeV (thick) and $1000$~GeV (dashed). The $\rm bbH$ mode becomes dominant for $\tan\beta \gtrsim 10$.\label{ggH-bbH-cxn}}
\end{figure}

Heavy neutral BSM Higgs bosons ($H$ and $A$) are primarily produced at the LHC via gluon fusion and bottom-quark-associated production ($\rm bbH/A$). As shown in Fig.~\ref{ggH-bbH-cxn}, gluon fusion proceeds through top- and bottom-quark loops and dominates at $\tan\beta \lesssim 10$, while $\rm bbH/A$ production scales with the bottom Yukawa coupling and becomes strongly enhanced at large $\tan\beta$.   
In the region $\tan\beta \gtrsim 30$, relevant to our parameter space, $\rm bbH/A$ exceeds gluon fusion by more than two orders of magnitude~\cite{Dumont:2014wha}, making it the dominant production mode.

\begin{figure}[t]
\centering
\includegraphics[width=0.95\linewidth]{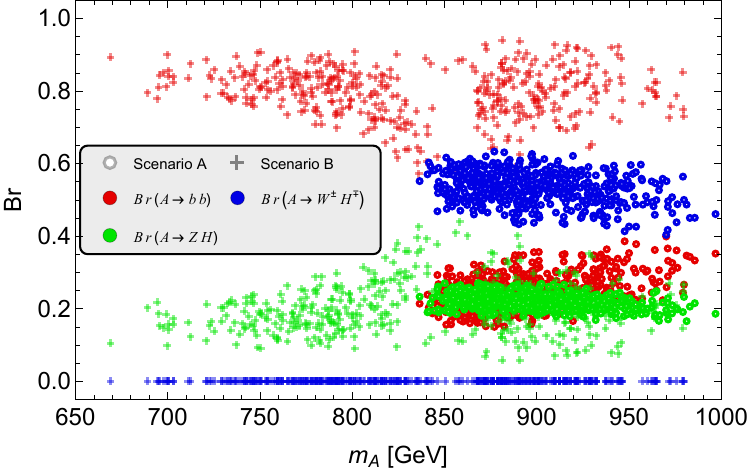}
\caption{
Branching ratios of the pseudoscalar $A$ for EWPT viable points with $\xi > 0.8$ in Scenario~A (circles) and Scenario~B (pluses). Several possible decay modes include $b\bar{b}$ (red), $ZH$ (green), and $W^\pm H^\mp$ (blue). In Scenario~B, $A \to b\bar{b}$ dominates, whereas in Scenario~A the decay $A \to W^\pm H^\mp$ is kinematically allowed and typically dominant, with $A \to b\bar{b}$ and $A \to ZH$ subleading.
\label{Br_HA}}
\end{figure}

Within the two viable scenarios, $H$ is either the lightest BSM Higgs boson (Scenario~B) or nearly degenerate with $H^\pm$ but lighter than $A$ (Scenario~A). Consequently, the decays $H \to AZ$ and $H \to W^\pm H^\mp$ are kinematically forbidden, and $H$ predominantly decays into $b\bar{b}$. 
The pseudoscalar $A$, typically the heaviest state in both scenarios, admits several decay channels, including $A\to ZH$, $A\to W^\pm H^\mp$ and $A \to b\bar{b}$, with branching ratios determined by the mass hierarchy.  
As illustrated in Fig.~\ref{Br_HA}, the decay $H/A \to b\bar b$ remains dominant across the viable EWPT parameter space, even when $t\bar t$ channels are kinematically accessible, making $\rm bbH/A$ production with $H/A \to b\bar b$ a particularly sensitive probe of the strong EWPT region.

A key distinction between the two scenarios lies in the decay $A \to W^\pm H^\mp$. This channel is kinematically allowed only in Scenario~A where the hierarchy $m_A > m_{H^\pm}+m_W$ is satisfied, but closed in Scenario~B where $m_A\simeq m_{H^\pm}$. 
Consequently, $A \to W^\pm H^\mp$ typically dominates in Scenario~A, whereas $A \to b\bar{b}$ becomes the main decay in Scenario~B due to both phase-space suppression of competing channels and the enhancement from large $\tan\beta$. 
Therefore, the decay $A \to W^\pm H^\mp$ provides a clean discriminator between the two scenarios and serves as a potential smoking-gun signature of the mass hierarchy responsible for strengthening the EWPT.

To assess collider sensitivity, we compute $bbH/A$ production cross sections using \texttt{HiggsTools}. Signal and background events are generated with \texttt{MadGraph5}~\cite{Alwall:2014hca}, showered with \texttt{Pythia8}~\cite{Sjostrand:2014zea}, and passed through \texttt{Delphes}~\cite{deFavereau:2013fsa} for detector simulation. 
Dominant SM backgrounds are generated at leading order and rescaled to next-to-leading-order (NLO) using appropriate $k$-factors (except for $t\bar t+X$). In the following, we present analyses for two representative channels: (i) $\rm bbH$ production with $H\rightarrow b\bar{b}$, and (ii) $\rm bbA$ Production with $A\to W^\pm H^\mp$, at the 13~TeV LHC with an integrated luminosity of $\mathcal{L}=300~{\rm fb}^{-1}$ and the 14~TeV HL-LHC with $\mathcal{L}=3~{\rm ab}^{-1}$. 
			
\subsection{Associated $\rm bbH$ production with $H\rightarrow b\bar{b}$}
\label{sec:4b_13}
			
The $\rm bbH$ production with $H \to b\bar b$ yields a characteristic four-$b$ final state. 
At large $\tan\beta$, both the production cross section and $\mathrm{Br}(H\to b\bar b)$ are enhanced, making this channel particularly sensitive to extended Higgs sectors relevant for a strong EWPT.
			
The dominant background arises from QCD multijet production ($pp\to bb\bar{b}\bar{b}$), while the $t\bar{t}$ contribution becomes negligible after event selection. 
Signal extraction relies on efficient $b$-tagging and invariant-mass reconstruction.
			
\begin{table*}[t]
\centering
\renewcommand{\tablename}{Table}
\caption{Results of the $H\to b\bar{b}$ analysis for BMPs at the LHC with $\sqrt{s}=13$ and $14~{\rm TeV}$. 
The table lists tree-level cross sections $\sigma_{\sqrt{s}}^{\rm bbH}$ for signals, cut efficiencies ($\epsilon$) for signal ($S$) and background ($B$), and the resulting significance $z \equiv S / \sqrt{S+B}$, where $S$ and $B$ denote event counts after selection. For the SM background, the NLO cross sections for $pp \to bb\bar{b}\bar{b}$ are 3556.8~pb at 13~TeV, obtained by rescaling the LO result with a $k$-factor of $1.72$, and 3771.6~pb at 14~TeV using $k=1.4$~\cite{Alwall:2014hca,Czakon:2015qwa}.}
\begin{tabular}{c|ccc|cccc|cccc}
\toprule
\multirow{2}{*}{BMP} & \multirow{2}{*}{$\tan\beta$} & \multirow{2}{*}{$m_H$ [GeV]} & \multirow{2}{*}{${\rm Br}_{H\rightarrow b\bar{b}}$} & \multicolumn{4}{c}{13 TeV} & \multicolumn{4}{c}{14 TeV} \\
\cmidrule(lr){5-8} \cmidrule(lr){9-12} & & & & $\sigma_{13}^{\rm bbH}$ [pb] & $\epsilon^S_{13}$ & $\epsilon^B_{13}$ & $z$ & $\sigma_{14}^{\rm bbH}$ [pb] & $\epsilon^S_{14}$ & $\epsilon^B_{14}$ & $z$ \\
\midrule
1 & 33 & 600 & 0.99 & 1.22 & 0.30 & 0.01 & 30  & 1.68 & 0.32 & 0.02 & 116 \\
2 & 32 & 705 & 0.99 & 0.53 & 0.30 & 0.007 & 17  & 0.76 & 0.32 & 0.01 & 68  \\
3 & 41 & 790 & 0.99 & 0.49 & 0.22 & 0.004 & 15  & 0.64 & 0.30 & 0.006 & 72  \\
\bottomrule
\end{tabular}
\label{res4bCombined}
\end{table*}

We perform Monte Carlo simulations at $\sqrt{s} = 13$~TeV and $14~{\rm TeV}$ for three benchmark points (BMPs) with increasing $m_H$, summarized in Table~\ref{res4bCombined}. Background events are generated with the basic parton-level cuts
\begin{description}
\item[(a)]  $p_T^b > 20$~GeV and $|\eta_b| < 2.4$ for all $b$-jets.
\end{description}
 
To exploit the kinematic features of the heavy-resonance, we apply the optimized cuts
\begin{description}
\item[(b)] $p_T^{b_1} > 0.23(0.24)~m_H$ for the leading $b$-jet, $p_T^{b_2} > 0.15(0.17)~m_H$ for the subleading $b$-jet
\item[(c)] $m_{bb} > 0.5(0.55)~m_H$ for invariant mass reconstructed from the two leading $b$-tagged jets
\end{description}
for 13 (14)~TeV. 
These cuts efficiently suppress the broad QCD background while preserving the boosted signal topology.
			
\begin{figure}[t]
\centering
\includegraphics[width=0.82\linewidth]{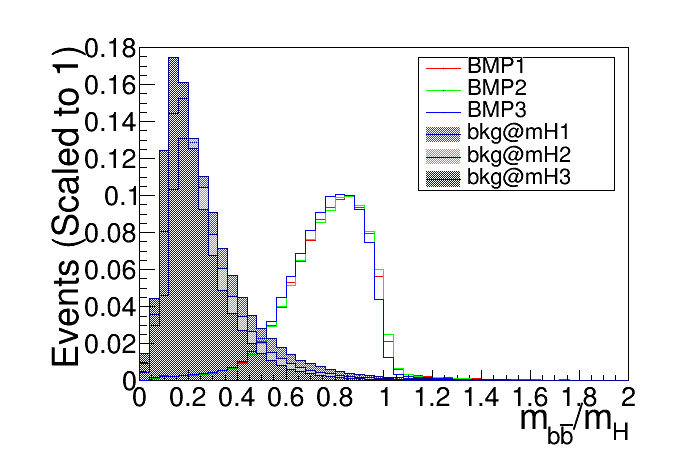}\\\vspace{-3mm}
\includegraphics[width=0.82\linewidth]{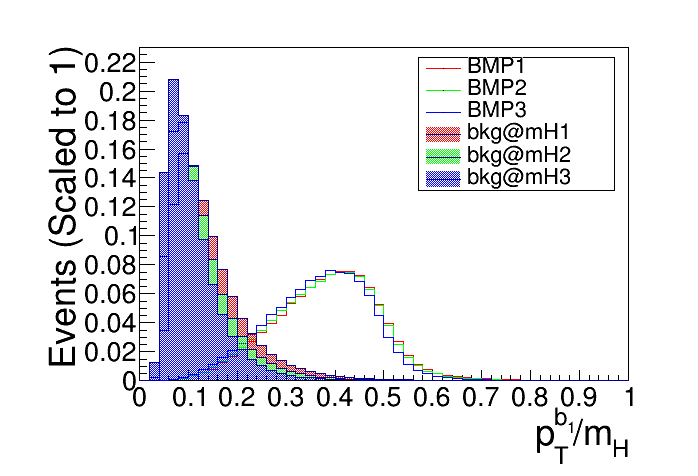}\\\vspace{-1mm}
\includegraphics[width=0.82\linewidth]{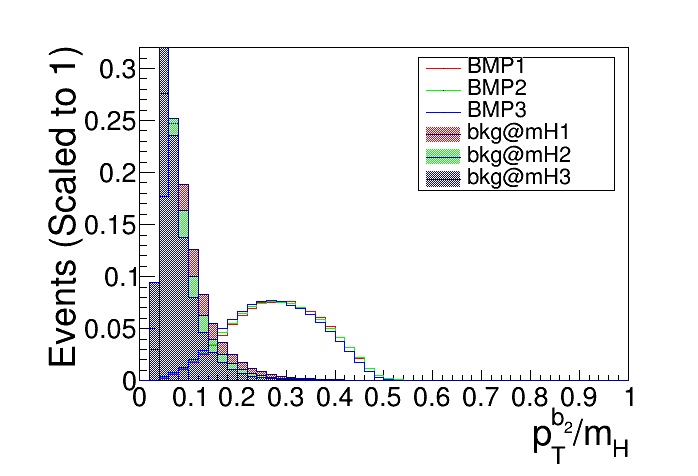}
\caption{Normalized distributions for the $H \to b\bar{b}$ channel at 13~TeV for the BMPs and the QCD background. (\textbf{Top}) Di-$b$-jet invariant mass, $m_{b\bar{b}}/m_H$, (\textbf{Middle}) transverse momentum of the leading $b$-jet, $p_T^{b_1}/m_H$,  and (\textbf{Bottom}) transverse momentum of the subleading $b$-jet, $p_T^{b_2}/m_H$. All distributions are normalized to unit area to allow direct comparison among BMPs with different $m_H$. Signal events exhibit a resonant peak at $m_{b\bar{b}}\simeq 1$ and harder $p_T$ spectra compared to the broad, featureless background, providing effective discriminants for optimized selection cuts. The absence of events in the low-$p_T$ region reflects the basic cut (a). \label{4bhisto13}}
\end{figure}
			
Fig.~\ref{4bhisto13} displays the normalized distributions of $m_{b\bar{b}}/m_H$, $p_T^{b_1}/m_H$, and $p_T^{b_2}/m_H$ for the signal and QCD background at 13~TeV prior to the optimized cuts. The di-$b$ invariant mass exhibits a clear resonant peak at $m_{b\bar{b}}\simeq m_H$ for each BMP, whereas the background a broad, smoothly falling spectrum. This clear resonance structure makes $m_{bb}$ the most powerful discriminant. In addition, the signal $p_T$ spectra of both the leading and subleading $b$-jets are significantly harder than those of the background, as expected from the decay of a heavy scalar. The leading $b$-jet provides the strongest separation, while the subleading jet offers complementary suppression of soft QCD events. These features directly motivate the optimized cuts (b) and (c).

The results after the full cut flow are summarized in Table~\ref{res4bCombined}. At the 13~TeV LHC, the signal efficiencies remain sizable ($\epsilon^S_{13} \simeq 0.2-0.3$), while the background efficiency is reduced to the percent level ($\epsilon^B_{13} \lesssim 0.01$). 
The resulting significances, defined as $z \equiv S/\sqrt{S+B}$, exceed the $5\sigma$ discovery threshold for all BMPs, reaching $z \simeq 15-30$ depending on $m_H$.
At the 14~TeV HL-LHC with $3~\mathrm{ab}^{-1}$, the slightly harder kinematics and larger luminosity substantially enhance the sensitivity, yielding $z \gtrsim 100$. 
The improvement approximately follows $\sqrt{\mathcal{L}}$ scaling and reflects both higher production rates and improved separation. Overall, this $H\to b\bar{b}$ channel provides robust discovery potential across all BMPs, particularly at the HL-LHC.

\subsection{Associated $\rm bbA$ Production with $A\to W^\pm H^\mp$}
\label{sec:WHc_13}

\begin{table*}[t]
\centering
\caption{Results of the $A \to W^\pm H^\mp$ analysis for the selected BMPs at the LHC with $\sqrt{s}=13$ and $14~{\rm TeV}$. The table lists the tree-level signal production cross sections $\sigma_{\sqrt{s}}^{\rm bbA}$, preselection cut efficiencies for signal $\epsilon^S_{\rm cut}$, BDT selection efficiencies $\epsilon_{\rm BDT}$ for signal ($S$) and background ($B$) at BDT, and the resulting significance $z$.}
\begin{tabular}{c|cccc|ccccc|ccccc}
\toprule
\multirow{2}{*}{BMP} & \multirow{2}{*}{$\tan\beta$} & \multirow{2}{*}{\shortstack{$m_A$ \\ $[\mathrm{GeV}]$}} & \multirow{2}{*}{\shortstack{$m_{H^\pm}$ \\ $[\mathrm{GeV}]$}} & \multirow{2}{*}{${\rm Br}_{A\rightarrow W^\pm H^\mp}$} & \multicolumn{5}{c}{13 TeV} & \multicolumn{5}{c}{14 TeV} \\
\cmidrule(lr){6-10} \cmidrule(lr){11-15} & & & & & $\sigma_{13}^{\rm bbA}$ [pb] & $\epsilon^{S}_{\rm{cut}}$ & $\epsilon^S_{\rm{BDT}}$ & $\epsilon^B_{\rm{BDT}}$ & $z$ & $\sigma_{14}^{\rm bbA}$ [pb] & $\epsilon^{S}_{\rm{cut}}$ & $\epsilon^S_{\rm{BDT}}$ & $\epsilon^B_{\rm{BDT}}$ & $z$ \\
\midrule
1 & 34  & 849 &  603 & 0.57 & 0.13 & 0.11  & 0.279  & 0.0024 & 1.20 &  0.18 & 0.11 & 0.292 & 0.0028 & 4.88 \\
2 & 34 & 875 &  649 & 0.52 & 0.10 & 0.11  & 0.338  & 0.0036 & 0.94 &  0.14 & 0.11 & 0.293 & 0.0027 & 3.73 \\
3 & 32  & 904 &  701 & 0.48 & 0.07 & 0.10  & 0.295  & 0.0030 & 0.57 &  0.10 & 0.10 & 0.323 & 0.003 & 2.7 \\
4 & 30  & 951 &  751 & 0.49 & 0.05 & 0.10  & 0.341  & 0.0036 & 0.47 &  0.07 & 0.10 & 0.313 & 0.0028 & 1.76 \\
5 & 36  & 998 &  801 & 0.43 & 0.05 & 0.10  & 0.375  & 0.0030 & 0.43 &  0.07 & 0.09 & 0.36 & 0.003 & 1.76 \\
\bottomrule
\end{tabular}
\label{resaw13}
\end{table*}

\begin{table}[t]
\centering
\caption{Background cross sections $\sigma_{\sqrt{s}}$ and corresponding cut efficiencies $\epsilon_{\rm{cut}}$ for the considered SM processes at the LHC with $\sqrt{s}=13$ and $14~{\rm TeV}$.}
\begin{tabular}{c|cc|cc}
\toprule
\multirow{2}{*}{SM backgroud} & \multicolumn{2}{c}{13 TeV} & \multicolumn{2}{c}{14 TeV} \\
\cmidrule(lr){2-3} \cmidrule(lr){4-5} & $\sigma^B_{13}$ [pb] & $\epsilon^B_{\rm{cut}}$ & $\sigma^B_{14}$ [pb] & $\epsilon^B_{\rm{cut}}$ \\
\midrule
$tt$   & 21.92 & 0.064 & 25.92 & 0.063 \\
$ttj$  & 13.36 & 0.064 & 16.07 &0.063 \\
$ttjj$ &  5.36 & 0.064 & 6.56 &0.063 \\
\bottomrule
\end{tabular}
\label{resaw_bg_combined}
\end{table}

In this subsection, we study the associated $\rm bbA$ production with $A \to W^\pm H^\mp$ in Scenario~A.  The subsequent decay $H^+ \to t \bar{b}$ dominates, primarily due to near-degeneracy $m_H \simeq m_{H^\pm}$ together with large $\tan\beta$, followed by $t \to W^+ b$.
Focusing on leptonic $W$ decays, $W^\pm \to \ell^\pm \nu$ ($\ell = e, \mu$), the resulting final state consists of four $b$-jets, two oppositely charged leptons, and missing transverse energy ($bb\bar{b}\bar{b} + \ell^+\ell^- + {E}^{\rm miss}_T$), 
providing distinct kinematic handles relative to the purely hadronic $bb\bar{b}\bar{b}$ final state.

The dominant SM backgrounds arise from $t \bar{t}$ and $t \bar{t}+$ jets production with $t\to W^+b$ and leptonic $W$ decays, while subleading contributions, such as multi-boson production, are neglected.  
Despite the large $t \bar{t}$ backgrounds, the rich kinematic structure of the signal, particularly reconstructed resonance masses and angular correlations, provides powerful observables for multivariate discrimination. 

To exploit these features, in this analysis we employ a two-stage strategy, a basic preselection followed by a multivariate analysis (MVA) implemented through boosted decision trees (BDTs). 
The preselection requires:
\begin{description}
\item[(a)] at least two $b$-tagged jets with $p_T > 20~{\rm GeV}$ and $|\eta| < 2.5$;
\item[(b)] at least two oppositely charged leptons ($e^+e^-, \mu^+\mu^-, e^\pm \mu^\mp$) with $p_T > 20~{\rm GeV}$ and $|\eta| < 2.4$.
\end{description}
The five BMPs from Scenario~A used in the analysis are listed in Table~\ref{resaw13}, with cross sections and efficiencies for the backgrounds summarized in Table~\ref{resaw_bg_combined}. 
The cut efficiencies, $\epsilon_{\rm cut}$, are computed after this preselection. 
These baseline selections mainly suppress trivial backgrounds and do not yet fully exploit the distinctive kinematic features of the signal. 

Although the signal and background processes share similar visible final states, a key difference is that the signal originates from heavy intermediate resonances, such as $H^\pm$ and $A$. We therefore exploit this feature by reconstructing the parent resonances via invariant mass variables, thereby enhancing discrimination between the signal and SM backgrounds. To this end, we reconstruct $H^\pm$ and $A$ states as follows:
\begin{description}
\item[(c)] reconstruct the $H^\pm$ candidate from the two leading $b$-jets and one lepton, computing its invariant mass  for each combination. Select the combination closest to $m_{H^\pm}$ and denote it $m_{H^\pm}^{\text{\rm rec}}$; 
\item[(d)] reconstruct the $A$ candidate using all visible objects and $E_T^{\rm miss}$, defining $m_A^{\text{rec}}$.
\end{description}

\begin{figure}[t]
\centering
\includegraphics[width=0.82\linewidth]{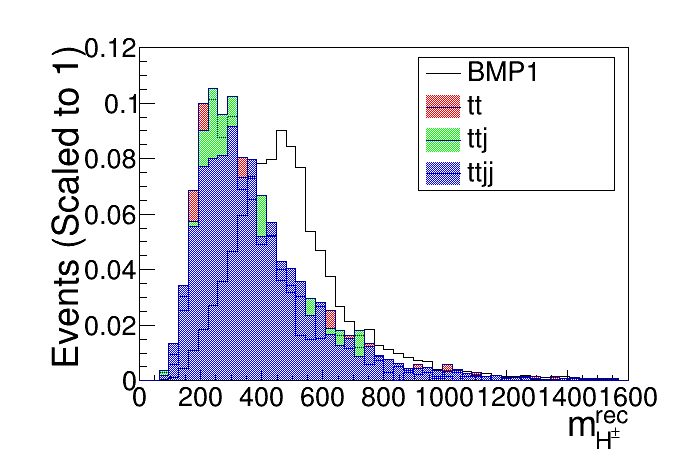}\\
\includegraphics[width=0.82\linewidth]{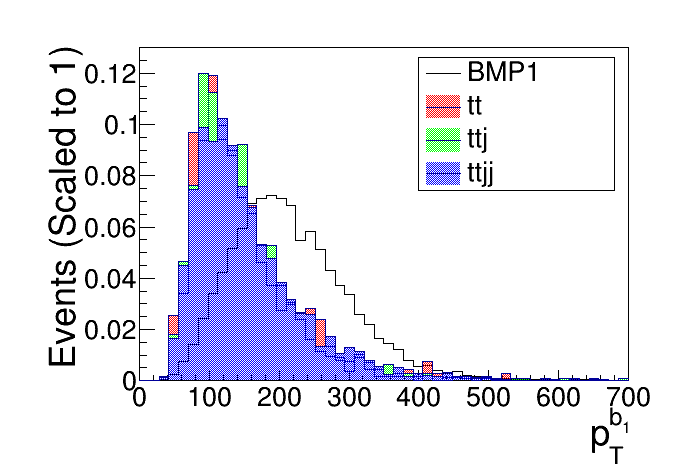}\\
\includegraphics[width=0.82\linewidth]{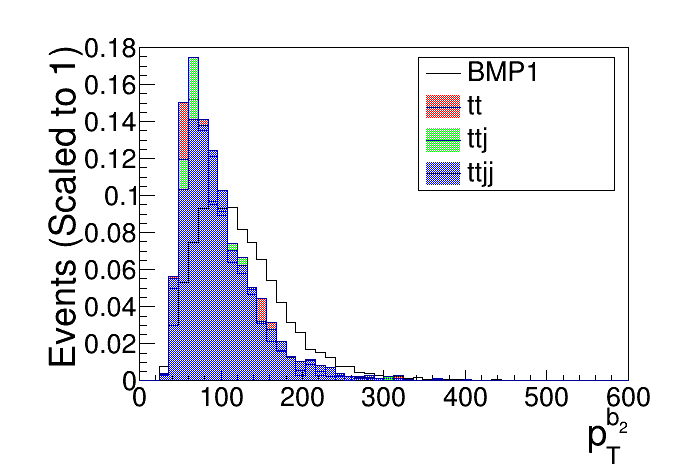}
\caption{Normalized kinematic distributions for BMP1 (signal) and the dominant SM backgrounds at $\sqrt{s}=13~\mathrm{TeV}$. (\textbf{Top}) Reconstructed invariant mass $m_{H^\pm}^{\rm rec}$. (\textbf{Middle}) Transverse momentum of the leading $b$-jet, $p_T^{b_1}$. (\textbf{Bottom}) Transverse momentum of the subleading $b$-jet, $p_T^{b_2}$. The signal shows a resonant structure in $m_{H^\pm}^{\rm rec}$ and moderately harder $p_T$ spectra than the $t\bar{t}$-associated backgrounds, though with noticeable overlap between signal and background distributions.\label{awhisto13}}
\end{figure}
						
Fig.~\ref{awhisto13} displays normalized kinematic distributions after preselection for BMP-1 at 13~TeV. 
The reconstructed charged Higgs mass $m_{H^\pm}^{\rm rec}$ shows a pronounced resonant peak close to the true mass for the signal, while the $t\bar t$-associated backgrounds exhibit broad, non-resonant spectra. The leading and subleading $b$-jet transverse momenta are also harder for signal events, particularly relative to the dominant $t\bar{t}jj$ background. However, compared to the $bb\bar{b}\bar{b}$ final state, the separation between signal and background is less pronounced, with substantial overlap persisting in all one-dimensional distributions. This reduced isolation motivates the search for additional discriminating observables and the use of a MVA to exploit correlations among variables for more efficient background suppression.

Given the close resemblance between the signal and the dominant $t\bar{t}$ background---both containing dileptons and multiple $b$-jets, we reconstruct $t\bar{t}$ systems for both background and hypothetical signal events in order to compute the angular observables $C_{\mathrm{hel}}$ and $C_{\mathrm{han}}$~\cite{Bernreuther:2015yna}. These observables probe spin correlations and CP structure of the intermediate state and thus can provide discrimination between SM-like $t\bar{t}$ production and signal process (i.e., $A \to t\bar{t}$) that mimic a $t\bar{t}$ topology~\cite{Arco:2025ydq}.

To perform this reconstruction, the four-momenta of the top quarks in the $t\bar{t}$ system are required. Following the method of Ref.~\cite{Sonnenschein:2006ud}, we solve for the momenta of the two neutrinos from the measured four-momenta of the two leptons and two $b$-jets, assuming the known masses of the $W^\pm$ boson and the top quark, and then compute the top momenta.
Physical solutions for the reconstructed neutrino momenta are obtained for most background events but for fewer than half of the signal events, reflecting the absence of a true $t\bar{t}$ topology in the latter. Events without physical solutions are assigned vanishing reconstructed top momenta. 

\begin{figure}[!t]
\centering
\includegraphics[width=0.72\linewidth]{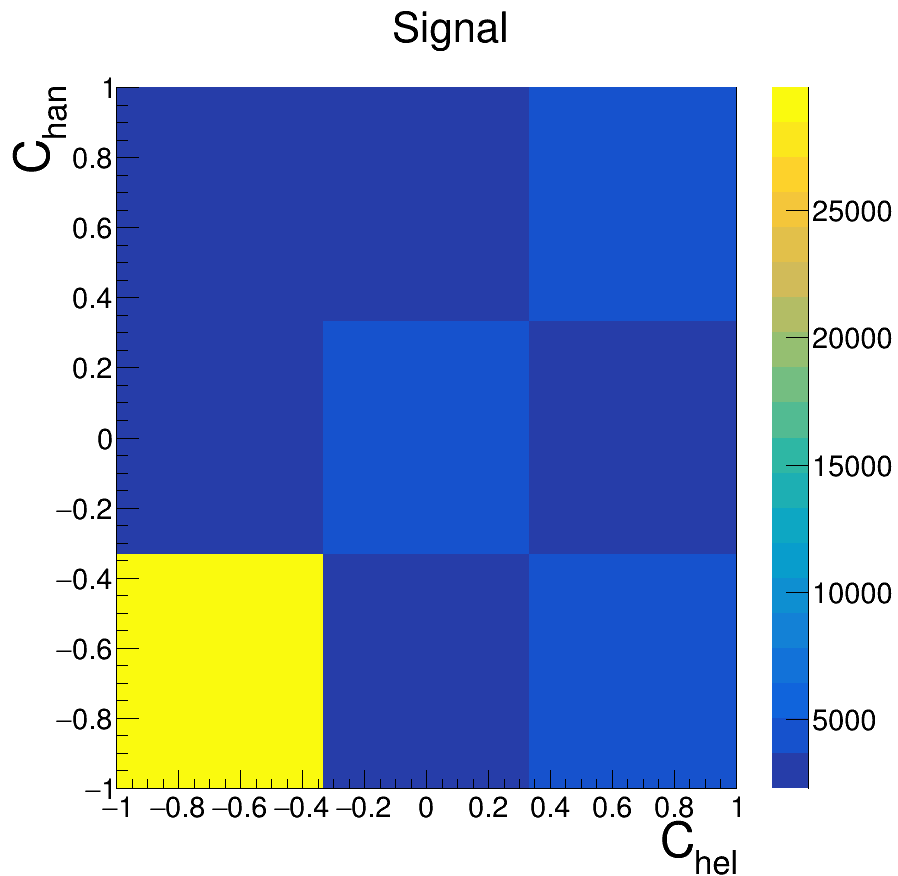}\\
\includegraphics[width=0.72\linewidth]{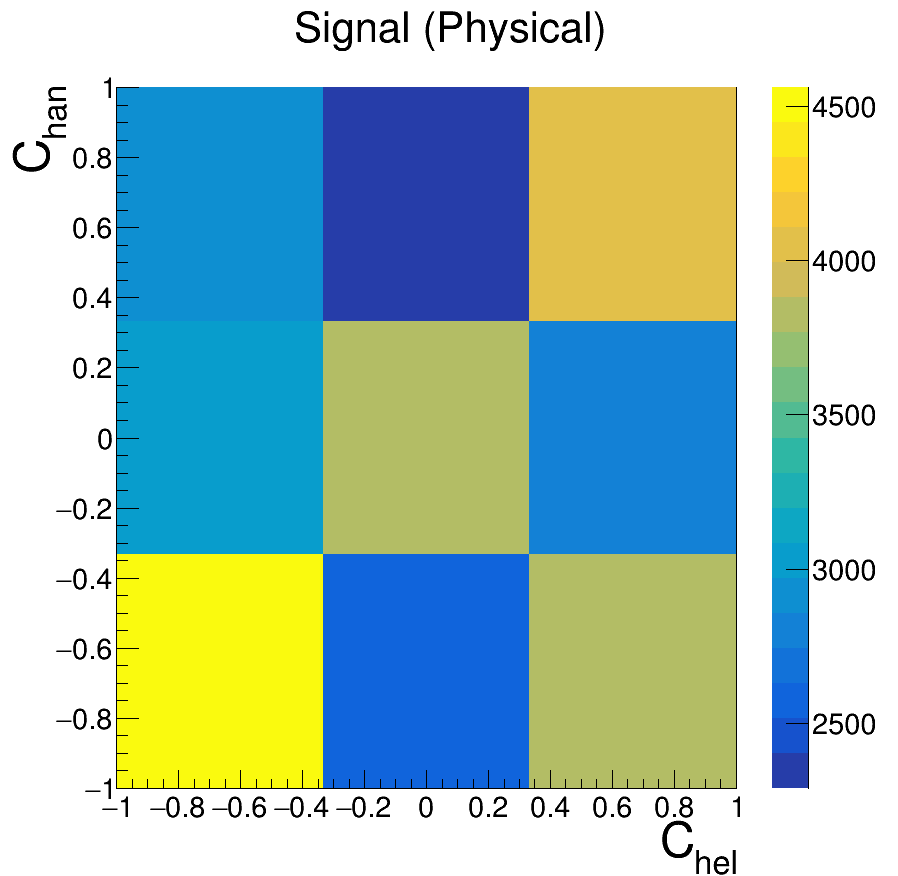}\\
\includegraphics[width=0.72\linewidth]{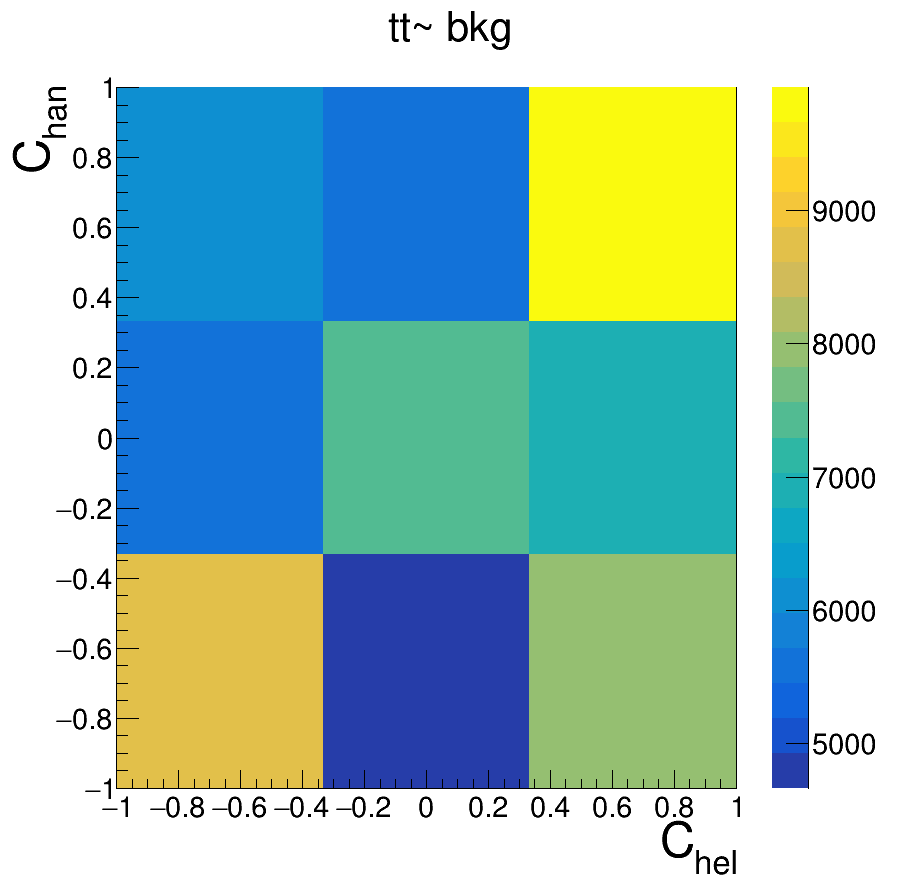}
\caption{Two-dimensional distribution of $(C_{\rm hel},, C_{\rm han})$ for (top) all signal events, (middle) signal events with physical $t\bar{t}$ reconstruction solutions, and (bottom) SM $t\bar{t}$ background events. Signal events, irrespective of whether nonphysical reconstruction solutions are excluded, cluster near $(-1,-1)$, while the background peaks around $(1,1)$; the smaller accumulation near $(-1,-1)$ arises from events without physical reconstruction solutions.
\label{CC-fig}}
\end{figure}

Using BMP-1 as a representative example, Fig.~\ref{CC-fig} shows that signal and background populate distinct regions in the $C_{\rm hel}$-$C_{\rm han}$ plane. 
Signal events cluster near $(-1,-1)$, whereas the $t\bar t+$jets background, closely resembling inclusive SM $t\bar t$ production, peaks around $(1,1)$. Notably, this separation persists even when restricting to signal events with physical solutions for the reconstructed neutrino momenta. Although the signal does not contain a genuine $t\bar t$ pair, this reconstruction nonetheless provides useful discrimination.
			
Building on these insights, we train BDTs using the Toolkit for MVA framework implemented in \texttt{ROOT}~\cite{Brun:1997pa}.\footnote{The Toolkit for MVA (TMVA)~\cite{TMVA:2007ngy} is a built-in ROOT package that provides a broad suite of machine-learning algorithms for high-energy physics applications.} The input variables include:
\begin{itemize}
\item kinematic observables: $p_T^{b_1}, p_T^{b_2}, p_T^{\ell_1}, p_T^{\ell_2}, E_T^{\rm miss}$;
\item reconstructed masses: $m_{H^\pm}^{\rm rec}$, $m_A^{\rm rec}$;
\item angular variables: $\Delta R$ between $b$-jets and leptons;
\item CP-sensitive observables: $C_{\text{hel}}, C_{\text{han}}$.
\end{itemize}
where $\Delta R\equiv \sqrt{(\Delta \phi)^2+(\Delta \eta)^2}$. 
\begin{figure}[t]
	\centering
	\includegraphics[width=0.95\linewidth]{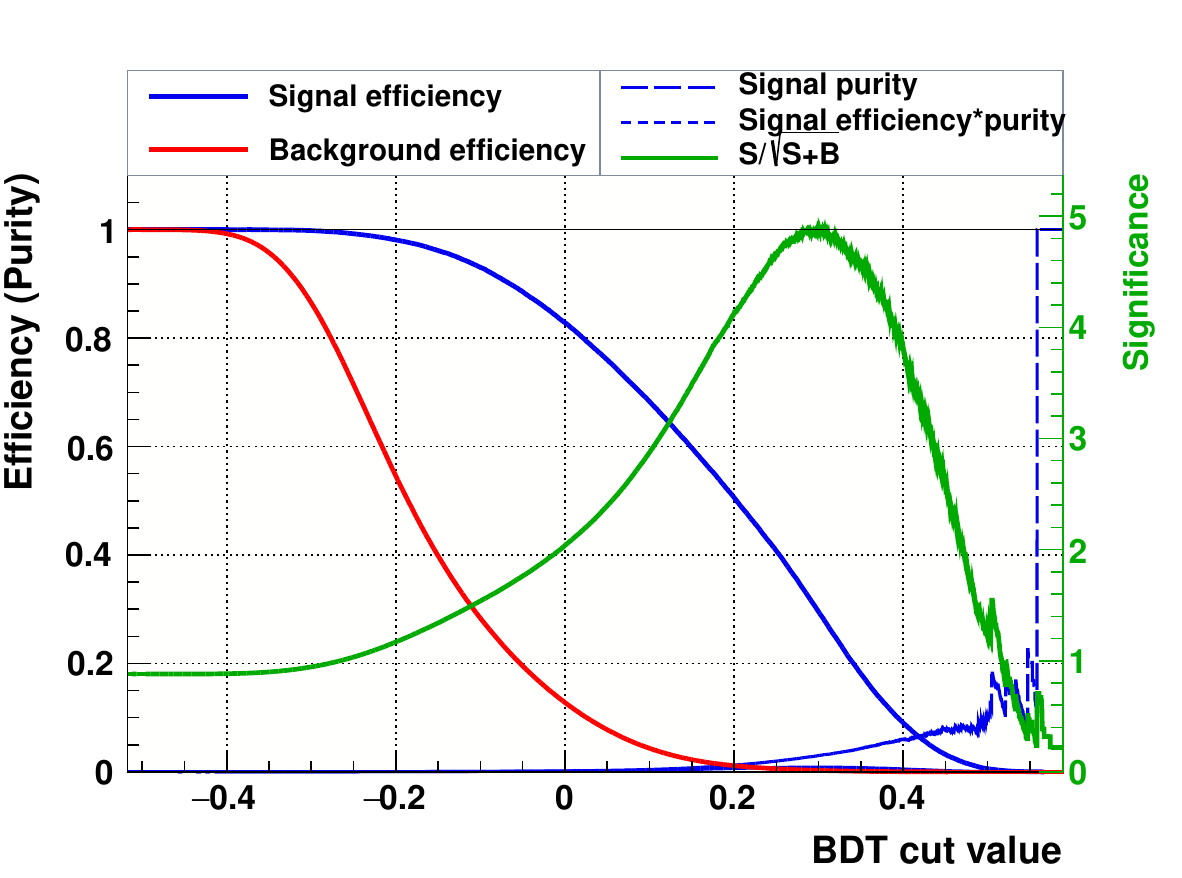}
	\caption{BDT signal significance, efficiencies and optimal cut value for BMP1 at 14 TeV LHC. x-axis is the cut value applied on the BDT out put and blue/red lines corresponding to the efficiencies for signal/background. And the green line represent the corresponding statistical significance calculated by the efficiencies and reaches a maximum of 4.9 at the cut value of 0.3012.  \label{BDT_info}}
\end{figure}

To quantify the sensitivity of the BDT analysis, we compute the statistical significance $z$ based on the expected signal and background yields after all selections. 
The signal yield is given by
\beq
S =\mathcal{L} \sigma^S  \epsilon^S_{\rm cut} \epsilon^S_{\rm BDT}
\eeq
where $\mathcal{L}$ is the integrated luminosity, $\sigma^S$ is the signal production cross section, and $\epsilon_{\rm cut}$ and $\epsilon_{\rm BDT}$ denote the efficiencies of the preselection and BDT stages, respectively.  
The total background yield $B$ is obtained by summing over all relevant SM background processes $B_i$, weighted by their respective cross sections $\sigma^{B_i}$ and cut efficiencies $\epsilon^{B_i}_{\rm cut}$. 
\beq
B  =\mathcal{L} \sum_{B_i} \left(\sigma^{B_i} \epsilon^{B_i}_{\rm cut}\right) \epsilon^B_{\rm BDT}
\eeq
where $\epsilon^B_{\rm BDT}$ is the BDT efficiency defined for the combined background sample, effectively corresponding to a weighted average over individual background components. 
With these definitions, the significance $z$ consistently incorporates the event reductions at both the preselection and BDT stages. Consequently, the final sensitivity depends sensitively on $\epsilon_{\rm BDT}$ for both signal and backgrounds. 

Figure~\ref{BDT_info} illustrates the BDT response for BMP~1 at 14 TeV. The BDT score serves as a multivariate discriminator, with higher values indicating events that are more signal-like according to the trained model. 
In particular, variables such as $p_T^{b_1}$ and $m_{H^\pm}^{\rm rec}$ provide strong discriminating power, while the full set of variables collectively determines the overall BDT performance.
As the BDT cut value increases, progressively larger fractions of both signal and background events are rejected, leading to decreasing BDT efficiencies for signals and backgrounds $\epsilon^S_{\rm BDT}$ and $\epsilon^B_{\rm BDT}$ to decrease. 
In the low BDT cut region, the background efficiency decreases more rapidly than that of the signal, improving the signal-to-background ratio and increasing the significance $z$. The significance $z$ reaches a maximum of 4.9 at the optimal cut value of 0.3. Beyond the optimal cut value, the signal efficiency is substantially reduced, causing the significance to decline. 

This behavior is generic across the BMPs, with BMPs~2–5 exhibiting similar trends. In all cases, the optimal BDT cut values cluster around 0.3, with differences in the final significance primarily driven by the signal event yields. As summarized in Table~\ref{resaw13}, BMP-1 and BMP-2, which feature relatively light $H^\pm$, achieve $z \simeq 3-4$, approaching the evidence level. 
For heavier $H^\pm$ (e.g., BMP-4 and BMP-5), the reduced production cross section lowers the sensitivity, although the BDT remains effective in suppressing backgrounds. While none of the individual BMPs reaches a $5\sigma$ discovery, the results clearly demonstrate the dependence of collider sensitivity on the scalar mass spectrum. 
In contrast, at 13~TeV the sensitivity remains limited, with $z \sim \mathcal{O}(1)$ due to the smaller signal cross sections and currently available luminosity, highlighting the importance of the HL-LHC for probing this scenario.

\begin{figure}[t]
\centering
\includegraphics[width=0.95\linewidth]{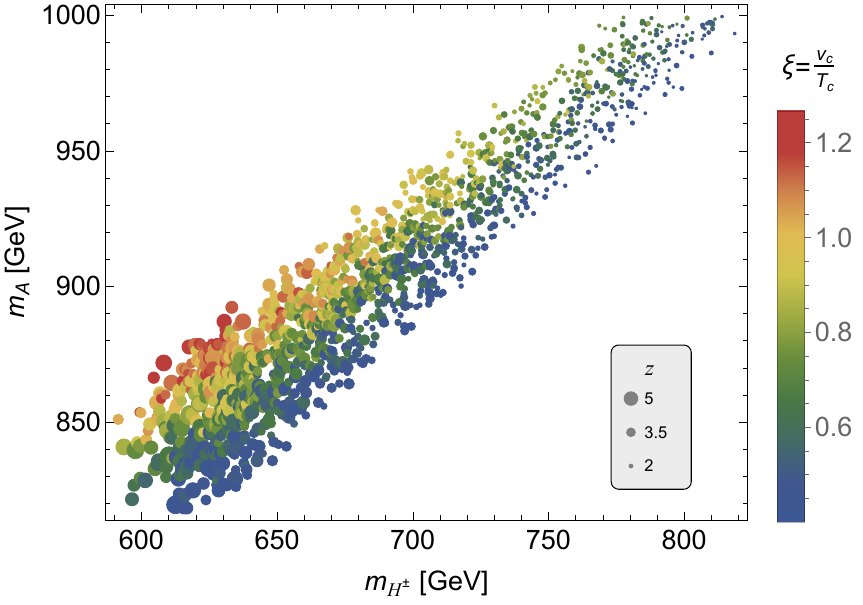}		
\caption{Signal significance and phase transition strength in Scenario~A at the 14 TeV HL-LHC with 3 ab$^{-1}$. The viable parameter space is displayed in the $(m_{H^\pm}, m_A)$ plane. Point colors encode the phase transition strength $\xi \equiv v_c/T_c$, while point sizes represent the signal significance $z$. The significance ranges from 1.4 to 6.8 (for comparison, it spans 0.4–2.2 at 13 TeV, not shown). Discovery-level sensitivity ($z \gtrsim 5$) is achieved in the lower-mass region around $m_{H^\pm} \simeq 600-650~{\rm GeV}$ and $m_A \simeq 850-900~{\rm GeV}$, which overlaps with points exhibiting stronger EWPTs, demonstrating that the HL-LHC can probe the parameter space most relevant for a strong EWPT.\label{aw-sig}}
\end{figure}

Extending the analysis to the full viable parameter space of Scenario~A, we find that at 13~TeV all points yield a signal significance $z<3$, remaining below the evidence threshold. In contrast, at the 14~TeV HL-LHC with an integrated luminosity of 3~ab$^{-1}$ (see Fig.~\ref{aw-sig}), the sensitivity improves substantially and can exceed the discovery threshold, reaching up to $z = 6.8$ in the most favorable regions of parameter space. The highest sensitivities occur for relatively light scalar masses, around $m_{H^\pm} \simeq 600-650~{\rm GeV}$ and $m_A \simeq 850-900~{\rm GeV}$, where larger production cross sections and more distinctive kinematic features enhance the signal selection efficiency.

We further examine the interplay between collider reach and the strength of the EWPT. Although there is no strict one-to-one correspondence between the signal significance $z$ and the phase transition strength $\xi$, points with larger $\xi$ tend to cluster in regions of higher collider sensitivity. This pattern indicates that the HL-LHC can effectively probe the parameter space most relevant for realizing a strong first-order EWPT in Scenario~A, providing a useful collider handle on the underlying scalar-sector dynamics.

\section{Conclusions and Outlook}
\label{sec:conc}
A first-order EWPT provides a compelling mechanism for generating the matter-antimatter asymmetry through EWBG. Since the SM predicts only a smooth crossover, realizing a first-order EWPT requires new physics. 2HDMs offer a minimal and well-motivated extension capable of accommodating such a phase transition while offering rich and testable collider phenomenology.

In this work, we studied the large-$\tan\beta$ regime of the flipped (Type-Y) 2HDM and identified two viable scenarios capable of generating a strong EWPT. Both scenarios feature a hierarchical Higgs spectrum with $m_A>m_H$, such that the CP-odd Higgs $A$ is the heaviest state. The key distinction lies in the mass of the charged Higgs $H^\pm$ relative to the neutral scalars. 
In Scenario~A, $H$ and $H^\pm$ are lighter and nearly degenerate, and the EWPT typically realized as a one-step transition occurring at relatively lower nucleation temperatures. 
In contrast, Scenario~B features a heavier $H^\pm$, close in mass to $A$, which not only gives rise to richer collider phenomenology but also admits both one-step and two-step EWPTs. 
Across all viable parameter regions, the nucleation condition is satisfied, ensuring successful completion of the phase transition and preventing false-vacuum trapping.

The distinct mass hierarchies in the two scenarios also shape the collider signatures. In both cases, the CP-even Higgs $H$ predominantly decays into $b\bar{b}$, motivating searches in bottom-associated production, $b\bar{b}H$ with $H \to b\bar{b}$. This channel is particularly effective in the large-$\tan\beta$ regime, where both production cross sections and branching fractions are enhanced, making it a highly sensitive probe of the parameter regions favored by a strong EWPT. 
With four tagged $b$-jets in the final state, the signal exhibits from a resonant invariant-mass peak and harder transverse-momentum spectra, which enable efficient suppression of QCD backgrounds.
Consequently, even a cut-based analysis yields very high sensitivity, with the significance readily exceeding the $5\sigma$ threshold. 
At the 14 TeV HL-LHC with an integrated luminosity of $3~\mathrm{ab}^{-1}$, the significance can reach $z \simeq 100$, demonstrating strong discovery potential across the viable parameter space.

While the $H\to b\bar{b}$ channel offers high-statistics sensitivity, the process $b\bar{b}A$ with $A \to W^\pm H^\mp$ provides a means to discriminate between the two EWPT scenarios, since this decay is kinematically forbidden in Scenario~B. Its observation would therefore uniquely identify Scenario~A and directly probe the BSM Higgs spectrum.
Although experimentally more challenging than the purely hadronic $bb\bar{b}\bar{b}$ final state, this channel provides additional kinematic handles, particularly through angular observables, which help improve background suppression. By applying optimized kinematic selections and a BDT-based MVA incorporating a broad set of observables, the HL-LHC can achieve $z \simeq 6.8$ in favorable regions of Scenario~A, . 
Notably, parameter regions with stronger phase transitions $\xi \gtrsim 1.2$ tend to coincide with enhanced collider sensitivity $z \geq 5$, indicating that HL-LHC searches can preferentially probe the parameter space most relevant for a strong first-order EWPT. 

Taken together, these complementary channels enables a systematic exploration of the BSM Higgs spectrum and its connection to early-Universe cosmology. The HL-LHC therefore provides a realistic opportunity not only to probe the scalar mass structures responsible for a strong first-order EWPT, but also to shed light on the underlying dynamics driving the EWPTs within the flipped 2HDM.

Building on these findings, several directions can further advance this framework. Experimentally, extending the analysis to additional production and decay channels, performing combined analyses of multiple final states, and adopting more sophisticated multivariate or machine-learning techniques would enhance the overall sensitivity and robustness of the collider program. On the theoretical side, incorporating explicit CP-violating sources and computing the resulting baryon asymmetry would enable a fully quantitative treatment of EWBG within the flipped 2HDM. 

\vspace{5mm}

\subsection*{Acknowledgments}
We are grateful to Kun Liu for insightful discussions and valuable suggestions regarding the collider search strategies. This work is supported by the National Key Research and Development Program of China (Grant No. 2021YFC2203002). Y. J. is also funded by the GuangDong Basic and Applied Basic Research Foundation (No. 2020A1515110150). W. S. is supported by the Natural Science Foundation of China (NSFC) under grant No. 12305115.

\bibliography{Refence}

@article{ATLAS:2012yve,
    author = "Aad, Georges and others",
    collaboration = "ATLAS",
    title = "{Observation of a new particle in the search for the Standard Model Higgs boson with the ATLAS detector at the LHC}",
    eprint = "1207.7214",
    archivePrefix = "arXiv",
    primaryClass = "hep-ex",
    reportNumber = "CERN-PH-EP-2012-218",
    doi = "10.1016/j.physletb.2012.08.020",
    journal = "Phys. Lett. B",
    volume = "716",
    pages = "1--29",
    year = "2012"
}

@article{CMS:2012qbp,
    author = "Chatrchyan, Serguei and others",
    collaboration = "CMS",
    title = "{Observation of a New Boson at a Mass of 125 GeV with the CMS Experiment at the LHC}",
    eprint = "1207.7235",
    archivePrefix = "arXiv",
    primaryClass = "hep-ex",
    reportNumber = "CMS-HIG-12-028, CERN-PH-EP-2012-220",
    doi = "10.1016/j.physletb.2012.08.021",
    journal = "Phys. Lett. B",
    volume = "716",
    pages = "30--61",
    year = "2012"
}

@article{WMAP:2012fli,
    author = "Bennett, C. L. and others",
    collaboration = "WMAP",
    title = "{Nine-Year Wilkinson Microwave Anisotropy Probe (WMAP) Observations: Final Maps and Results}",
    eprint = "1212.5225",
    archivePrefix = "arXiv",
    primaryClass = "astro-ph.CO",
    doi = "10.1088/0067-0049/208/2/20",
    journal = "Astrophys. J. Suppl.",
    volume = "208",
    pages = "20",
    year = "2013"
}

@article{Morrissey:2012db,
    author = "Morrissey, David E. and Ramsey-Musolf, Michael J.",
    title = "{Electroweak baryogenesis}",
    eprint = "1206.2942",
    archivePrefix = "arXiv",
    primaryClass = "hep-ph",
    reportNumber = "NPAC-12-08",
    doi = "10.1088/1367-2630/14/12/125003",
    journal = "New J. Phys.",
    volume = "14",
    pages = "125003",
    year = "2012"
}

@article{Sakharov:1967dj,
    author = "Sakharov, A. D.",
    title = "{Violation of CP Invariance, C asymmetry, and baryon asymmetry of the universe}",
    doi = "10.1070/PU1991v034n05ABEH002497",
    journal = "Pisma Zh. Eksp. Teor. Fiz.",
    volume = "5",
    pages = "32--35",
    year = "1967"
}

@article{Moore:1998swa,
    author = "Moore, Guy D.",
    title = "{Measuring the broken phase sphaleron rate nonperturbatively}",
    eprint = "hep-ph/9805264",
    archivePrefix = "arXiv",
    reportNumber = "MCGILL-98-7",
    doi = "10.1103/PhysRevD.59.014503",
    journal = "Phys. Rev. D",
    volume = "59",
    pages = "014503",
    year = "1999"
}

@article{LISA:2017pwj,
    author = "Amaro-Seoane, Pau and others",
    collaboration = "LISA",
    title = "{Laser Interferometer Space Antenna}",
    eprint = "1702.00786",
    archivePrefix = "arXiv",
    primaryClass = "astro-ph.IM",
    month = "2",
    year = "2017"
}

@article{TianQin:2015yph,
    author = "Luo, Jun and others",
    collaboration = "TianQin",
    title = "{TianQin: a space-borne gravitational wave detector}",
    eprint = "1512.02076",
    archivePrefix = "arXiv",
    primaryClass = "astro-ph.IM",
    doi = "10.1088/0264-9381/33/3/035010",
    journal = "Class. Quant. Grav.",
    volume = "33",
    number = "3",
    pages = "035010",
    year = "2016"
}

@article{Caprini:2019egz,
    author = "Caprini, Chiara and others",
    title = "{Detecting gravitational waves from cosmological phase transitions with LISA: an update}",
    eprint = "1910.13125",
    archivePrefix = "arXiv",
    primaryClass = "astro-ph.CO",
    reportNumber = "DESY-19-159, IPPP/19/27, HIP-2019-14/TH, MITP/19-066, IFT-UAM/CSIC-19-139",
    doi = "10.1088/1475-7516/2020/03/024",
    journal = "JCAP",
    volume = "03",
    pages = "024",
    year = "2020"
}

@article{Hu:2017mde,
    author = "Hu, Wen-Rui and Wu, Yue-Liang",
    title = "{The Taiji Program in Space for gravitational wave physics and the nature of gravity}",
    doi = "10.1093/nsr/nwx116",
    journal = "Natl. Sci. Rev.",
    volume = "4",
    number = "5",
    pages = "685--686",
    year = "2017"
}

@article{Kajantie:1996mn,
    author = "Kajantie, K. and Laine, M. and Rummukainen, K. and Shaposhnikov, Mikhail E.",
    title = "{Is there a~ hot electroweak phase transition at $m_H \gtrsim m_W$?}",
    eprint = "hep-ph/9605288",
    archivePrefix = "arXiv",
    reportNumber = "CERN-TH-96-126, HD-THEP-96-15, IUHET-333",
    doi = "10.1103/PhysRevLett.77.2887",
    journal = "Phys. Rev. Lett.",
    volume = "77",
    pages = "2887--2890",
    year = "1996"
}

@article{Kajantie:1996qd,
    author = "Kajantie, K. and Laine, M. and Rummukainen, K. and Shaposhnikov, Mikhail E.",
    title = "{A Nonperturbative analysis of the finite T phase transition in SU(2) x U(1) electroweak theory}",
    eprint = "hep-lat/9612006",
    archivePrefix = "arXiv",
    reportNumber = "BI-TP-96-54, CERN-TH-96-334A, HD-THEP-96-48",
    doi = "10.1016/S0550-3213(97)00164-8",
    journal = "Nucl. Phys. B",
    volume = "493",
    pages = "413--438",
    year = "1997"
}

@article{Csikor:1998eu,
    author = "Csikor, F. and Fodor, Z. and Heitger, J.",
    title = "{Endpoint of the hot electroweak phase transition}",
    eprint = "hep-ph/9809291",
    archivePrefix = "arXiv",
    reportNumber = "ITP-BUDAPEST-541, KEK-TH-580, KEK-PREPRINT-98-160, MS-TPI-98-16",
    doi = "10.1103/PhysRevLett.82.21",
    journal = "Phys. Rev. Lett.",
    volume = "82",
    pages = "21--24",
    year = "1999"
}

@article{Aoki:1999fi,
    author = "Aoki, Y. and Csikor, F. and Fodor, Z. and Ukawa, A.",
    title = "{The Endpoint of the first order phase transition of the SU(2) gauge Higgs model on a four-dimensional isotropic lattice}",
    eprint = "hep-lat/9901021",
    archivePrefix = "arXiv",
    reportNumber = "ITP-BUDAPEST-547, UTCCP-P-60, UTHEP-397",
    doi = "10.1103/PhysRevD.60.013001",
    journal = "Phys. Rev. D",
    volume = "60",
    pages = "013001",
    year = "1999"
}

@article{Branco:2011iw,
    author = "Branco, G. C. and Ferreira, P. M. and Lavoura, L. and Rebelo, M. N. and Sher, Marc and Silva, Joao P.",
    title = "{Theory and phenomenology of two-Higgs-doublet models}",
    eprint = "1106.0034",
    archivePrefix = "arXiv",
    primaryClass = "hep-ph",
    doi = "10.1016/j.physrep.2012.02.002",
    journal = "Phys. Rept.",
    volume = "516",
    pages = "1--102",
    year = "2012"
}

@article{Cline:1996mga,
    author = "Cline, James M. and Lemieux, Pierre-Anthony",
    title = "{Electroweak phase transition in two Higgs doublet models}",
    eprint = "hep-ph/9609240",
    archivePrefix = "arXiv",
    reportNumber = "MCGILL-96-16",
    doi = "10.1103/PhysRevD.55.3873",
    journal = "Phys. Rev. D",
    volume = "55",
    pages = "3873--3881",
    year = "1997"
}

@article{Fromme:2006cm,
    author = "Fromme, Lars and Huber, Stephan J. and Seniuch, Michael",
    title = "{Baryogenesis in the two-Higgs doublet model}",
    eprint = "hep-ph/0605242",
    archivePrefix = "arXiv",
    reportNumber = "CERN-PH-TH-2006-094, BI-TP-2006-18",
    doi = "10.1088/1126-6708/2006/11/038",
    journal = "JHEP",
    volume = "11",
    pages = "038",
    year = "2006"
}

@article{Cline:2011mm,
    author = "Cline, James M. and Kainulainen, Kimmo and Trott, Michael",
    title = "{Electroweak Baryogenesis in Two Higgs Doublet Models and B meson anomalies}",
    eprint = "1107.3559",
    archivePrefix = "arXiv",
    primaryClass = "hep-ph",
    doi = "10.1007/JHEP11(2011)089",
    journal = "JHEP",
    volume = "11",
    pages = "089",
    year = "2011"
}

@article{Dorsch:2014qja,
    author = "Dorsch, G. C. and Huber, S. J. and Mimasu, K. and No, J. M.",
    title = "{Echoes of the Electroweak Phase Transition: Discovering a second Higgs doublet through $A_0 \rightarrow ZH_0$}",
    eprint = "1405.5537",
    archivePrefix = "arXiv",
    primaryClass = "hep-ph",
    doi = "10.1103/PhysRevLett.113.211802",
    journal = "Phys. Rev. Lett.",
    volume = "113",
    number = "21",
    pages = "211802",
    year = "2014"
}

@article{Dorsch:2016nrg,
    author = "Dorsch, G. C. and Huber, S. J. and Konstandin, T. and No, J. M.",
    title = "{A Second Higgs Doublet in the Early Universe: Baryogenesis and Gravitational Waves}",
    eprint = "1611.05874",
    archivePrefix = "arXiv",
    primaryClass = "hep-ph",
    reportNumber = "DESY-16-213",
    doi = "10.1088/1475-7516/2017/05/052",
    journal = "JCAP",
    volume = "05",
    pages = "052",
    year = "2017"
}

@article{Basler:2016obg,
    author = "Basler, P. and Krause, M. and Muhlleitner, M. and Wittbrodt, J. and Wlotzka, A.",
    title = "{Strong First Order Electroweak Phase Transition in the CP-Conserving 2HDM Revisited}",
    eprint = "1612.04086",
    archivePrefix = "arXiv",
    primaryClass = "hep-ph",
    doi = "10.1007/JHEP02(2017)121",
    journal = "JHEP",
    volume = "02",
    pages = "121",
    year = "2017"
}

@article{Haarr:2016qzq,
    author = "Haarr, Anders and Kvellestad, Anders and Petersen, Troels C.",
    title = "{Disfavouring Electroweak Baryogenesis and a hidden Higgs in a CP-violating Two-Higgs-Doublet Model}",
    eprint = "1611.05757",
    archivePrefix = "arXiv",
    primaryClass = "hep-ph",
    reportNumber = "NORDITA-2016-121",
    month = "11",
    year = "2016"
}

@article{Basler:2017uxn,
    author = {Basler, Philipp and M{\"u}hlleitner, Margarete and Wittbrodt, Jonas},
    title = "{The CP-Violating 2HDM in Light of a Strong First Order Electroweak Phase Transition and Implications for Higgs Pair Production}",
    eprint = "1711.04097",
    archivePrefix = "arXiv",
    primaryClass = "hep-ph",
    reportNumber = "DESY-17-174, KA-TP-39-2017",
    doi = "10.1007/JHEP03(2018)061",
    journal = "JHEP",
    volume = "03",
    pages = "061",
    year = "2018"
}

@article{Bernon:2017jgv,
    author = "Bernon, J{\'e}r{\'e}my and Bian, Ligong and Jiang, Yun",
    title = "{A new insight into the phase transition in the early Universe with two Higgs doublets}",
    eprint = "1712.08430",
    archivePrefix = "arXiv",
    primaryClass = "hep-ph",
    doi = "10.1007/JHEP05(2018)151",
    journal = "JHEP",
    volume = "05",
    pages = "151",
    year = "2018"
}

@article{Parwani:1991gq,
    author = "Parwani, Rajesh R.",
    title = "{Resummation in a hot scalar field theory}",
    eprint = "hep-ph/9204216",
    archivePrefix = "arXiv",
    reportNumber = "ITP-SB-91-64",
    doi = "10.1103/PhysRevD.45.4695",
    journal = "Phys. Rev. D",
    volume = "45",
    pages = "4695",
    year = "1992",
    note = "[Erratum: Phys.Rev.D 48, 5965 (1993)]"
}

@article{Arnold:1992rz,
    author = "Arnold, Peter Brockway and Espinosa, Olivier",
    title = "{The Effective potential and first order phase transitions: Beyond leading-order}",
    eprint = "hep-ph/9212235",
    archivePrefix = "arXiv",
    reportNumber = "UW-PT-92-18, USM-TH-60",
    doi = "10.1103/PhysRevD.47.3546",
    journal = "Phys. Rev. D",
    volume = "47",
    pages = "3546",
    year = "1993",
    note = "[Erratum: Phys.Rev.D 50, 6662 (1994)]"
}

@article{ATLAS:2018rnh,
    author = "Aaboud, Morad and others",
    collaboration = "ATLAS",
    title = "{Search for pair production of Higgs bosons in the $b\bar{b}b\bar{b}$ final state using proton-proton collisions at $\sqrt{s} = 13$ TeV with the ATLAS detector}",
    eprint = "1804.06174",
    archivePrefix = "arXiv",
    primaryClass = "hep-ex",
    reportNumber = "CERN-EP-2018-029",
    doi = "10.1007/JHEP01(2019)030",
    journal = "JHEP",
    volume = "01",
    pages = "030",
    year = "2019"
}

@article{Kling:2020hmi,
    author = "Kling, Felix and Su, Shufang and Su, Wei",
    title = "{2HDM Neutral Scalars under the LHC}",
    eprint = "2004.04172",
    archivePrefix = "arXiv",
    primaryClass = "hep-ph",
    doi = "10.1007/JHEP06(2020)163",
    journal = "JHEP",
    volume = "06",
    pages = "163",
    year = "2020"
}

@article{Goncalves:2022wbp,
    author = "Gon{\c{c}}alves, Dorival and Kaladharan, Ajay and Wu, Yongcheng",
    title = "{Resonant top pair searches at the LHC: A window to the electroweak phase transition}",
    eprint = "2206.08381",
    archivePrefix = "arXiv",
    primaryClass = "hep-ph",
    doi = "10.1103/PhysRevD.107.075040",
    journal = "Phys. Rev. D",
    volume = "107",
    number = "7",
    pages = "075040",
    year = "2023"
}

@article{Haller:2018nnx,
    author = {Haller, Johannes and Hoecker, Andreas and Kogler, Roman and M{\"o}nig, Klaus and Peiffer, Thomas and Stelzer, J{\"o}rg},
    title = "{Update of the global electroweak fit and constraints on two-Higgs-doublet models}",
    eprint = "1803.01853",
    archivePrefix = "arXiv",
    primaryClass = "hep-ph",
    doi = "10.1140/epjc/s10052-018-6131-3",
    journal = "Eur. Phys. J. C",
    volume = "78",
    number = "8",
    pages = "675",
    year = "2018"
}

@article{Dubinin:1998nt,
    author = "Dubinin, M. N. and Semenov, A. V.",
    title = "{Triple and quartic interactions of Higgs bosons in the general two Higgs doublet model}",
    eprint = "hep-ph/9812246",
    archivePrefix = "arXiv",
    reportNumber = "SNUTP-98-140",
    month = "12",
    year = "1998"
}

@article{Bernon:2015qea,
    author = "Bernon, J{\'e}r{\'e}my and Gunion, John F. and Haber, Howard E. and Jiang, Yun and Kraml, Sabine",
    title = "{Scrutinizing the alignment limit in two-Higgs-doublet models: m$_h$=125  GeV}",
    eprint = "1507.00933",
    archivePrefix = "arXiv",
    primaryClass = "hep-ph",
    doi = "10.1103/PhysRevD.92.075004",
    journal = "Phys. Rev. D",
    volume = "92",
    number = "7",
    pages = "075004",
    year = "2015"
}

@article{Bernon:2015wef,
    author = "Bernon, J{\'e}r{\'e}my and Gunion, John F. and Haber, Howard E. and Jiang, Yun and Kraml, Sabine",
    title = "{Scrutinizing the alignment limit in two-Higgs-doublet models. II. m$_H$=125  GeV}",
    eprint = "1511.03682",
    archivePrefix = "arXiv",
    primaryClass = "hep-ph",
    doi = "10.1103/PhysRevD.93.035027",
    journal = "Phys. Rev. D",
    volume = "93",
    number = "3",
    pages = "035027",
    year = "2016"
}

@article{Dorsch:2016tab,
    author = "Dorsch, G. C. and Huber, S. J. and Mimasu, K. and No, J. M.",
    title = "{Hierarchical versus degenerate 2HDM: The LHC run 1 legacy at the onset of run 2}",
    eprint = "1601.04545",
    archivePrefix = "arXiv",
    primaryClass = "hep-ph",
    reportNumber = "DESY-15-240",
    doi = "10.1103/PhysRevD.93.115033",
    journal = "Phys. Rev. D",
    volume = "93",
    number = "11",
    pages = "115033",
    year = "2016"
}

@article{Karmakar:2019vnq,
    author = "Karmakar, Siddhartha and Rakshit, Subhendu",
    title = "{Relaxed constraints on the heavy scalar masses in 2HDM}",
    eprint = "1901.11361",
    archivePrefix = "arXiv",
    primaryClass = "hep-ph",
    doi = "10.1103/PhysRevD.100.055016",
    journal = "Phys. Rev. D",
    volume = "100",
    number = "5",
    pages = "055016",
    year = "2019"
}

@article{Li:2024kpd,
    author = "Li, Juxiang and Song, Huayang and Su, Shufang and Su, Wei",
    title = "{Charged Higgs search in 2HDM}",
    eprint = "2412.04572",
    archivePrefix = "arXiv",
    primaryClass = "hep-ph",
    doi = "10.1007/JHEP05(2025)063",
    journal = "JHEP",
    volume = "05",
    pages = "063",
    year = "2025"
}

@article{Atkinson:2021eox,
    author = "Atkinson, Oliver and Black, Matthew and Lenz, Alexander and Rusov, Aleksey and Wynne, James",
    title = "{Cornering the Two Higgs Doublet Model Type II}",
    eprint = "2107.05650",
    archivePrefix = "arXiv",
    primaryClass = "hep-ph",
    reportNumber = "SI-HEP-2021-20, SFB-257-P3H-21-049",
    doi = "10.1007/JHEP04(2022)172",
    journal = "JHEP",
    volume = "04",
    pages = "172",
    year = "2022"
}

@article{Misiak:2020vlo,
    author = "Misiak, M. and Rehman, Abdur and Steinhauser, Matthias",
    title = "{Towards $ \overline{B}\to {X}_s\gamma $ at the NNLO in QCD without interpolation in m$_{c}$}",
    eprint = "2002.01548",
    archivePrefix = "arXiv",
    primaryClass = "hep-ph",
    reportNumber = "TTP20-001, P3H-20-005, IFT-01/2020",
    doi = "10.1007/JHEP06(2020)175",
    journal = "JHEP",
    volume = "06",
    pages = "175",
    year = "2020"
}

@article{Misiak:2017bgg,
    author = "Misiak, Mikolaj and Steinhauser, Matthias",
    title = "{Weak radiative decays of the B meson and bounds on $M_{H^\pm }$ in the Two-Higgs-Doublet Model}",
    eprint = "1702.04571",
    archivePrefix = "arXiv",
    primaryClass = "hep-ph",
    reportNumber = "TTP17-004, IFT-1-2017",
    doi = "10.1140/epjc/s10052-017-4776-y",
    journal = "Eur. Phys. J. C",
    volume = "77",
    number = "3",
    pages = "201",
    year = "2017"
}

@article{Arbey:2017gmh,
    author = "Arbey, A. and Mahmoudi, F. and Stal, O. and Stefaniak, T.",
    title = "{Status of the Charged Higgs Boson in Two Higgs Doublet Models}",
    eprint = "1706.07414",
    archivePrefix = "arXiv",
    primaryClass = "hep-ph",
    reportNumber = "CERN-TH-2017-137, SCIPP-17-07",
    doi = "10.1140/epjc/s10052-018-5651-1",
    journal = "Eur. Phys. J. C",
    volume = "78",
    number = "3",
    pages = "182",
    year = "2018"
}

@article{BaBar:2012obs,
    author = "Lees, J. P. and others",
    collaboration = "BaBar",
    title = "{Evidence for an excess of $\bar{B} \to D^{(*)} \tau^-\bar{\nu}_\tau$ decays}",
    eprint = "1205.5442",
    archivePrefix = "arXiv",
    primaryClass = "hep-ex",
    reportNumber = "BABAR-PUB-12-012, SLAC-PUB-15028",
    doi = "10.1103/PhysRevLett.109.101802",
    journal = "Phys. Rev. Lett.",
    volume = "109",
    pages = "101802",
    year = "2012"
}

@article{Eriksson:2009ws,
    author = "Eriksson, David and Rathsman, Johan and Stal, Oscar",
    title = "{2HDMC: Two-Higgs-Doublet Model Calculator Physics and Manual}",
    eprint = "0902.0851",
    archivePrefix = "arXiv",
    primaryClass = "hep-ph",
    doi = "10.1016/j.cpc.2009.09.011",
    journal = "Comput. Phys. Commun.",
    volume = "181",
    pages = "189--205",
    year = "2010"
}

@article{Bahl:2022igd,
    author = {Bahl, Henning and Biek{\"o}tter, Thomas and Heinemeyer, Sven and Li, Cheng and Paasch, Steven and Weiglein, Georg and Wittbrodt, Jonas},
    title = "{HiggsTools: BSM scalar phenomenology with new versions of HiggsBounds and HiggsSignals}",
    eprint = "2210.09332",
    archivePrefix = "arXiv",
    primaryClass = "hep-ph",
    doi = "10.1016/j.cpc.2023.108803",
    journal = "Comput. Phys. Commun.",
    volume = "291",
    pages = "108803",
    year = "2023"
}

@article{ATLAS:2019tpq,
    author = "Aad, Georges and others",
    collaboration = "ATLAS",
    title = "{Search for heavy neutral Higgs bosons produced in association with $b$-quarks and decaying into $b$-quarks at $\sqrt{s}=13$ TeV with the ATLAS detector}",
    eprint = "1907.02749",
    archivePrefix = "arXiv",
    primaryClass = "hep-ex",
    reportNumber = "CERN-EP-2019-092",
    doi = "10.1103/PhysRevD.102.032004",
    journal = "Phys. Rev. D",
    volume = "102",
    number = "3",
    pages = "032004",
    year = "2020"
}

@article{CMS:2018hir,
    author = "Sirunyan, Albert M and others",
    collaboration = "CMS",
    title = "{Search for beyond the standard model Higgs bosons decaying into a $\mathrm{b\overline{b}}$ pair in pp collisions at $\sqrt{s} =$ 13 TeV}",
    eprint = "1805.12191",
    archivePrefix = "arXiv",
    primaryClass = "hep-ex",
    reportNumber = "CMS-HIG-16-018, CERN-EP-2018-124",
    doi = "10.1007/JHEP08(2018)113",
    journal = "JHEP",
    volume = "08",
    pages = "113",
    year = "2018"
}

@article{ATLAS:2018oht,
    author = "Aaboud, Morad and others",
    collaboration = "ATLAS",
    title = "{Search for a heavy Higgs boson decaying into a $Z$ boson and another heavy Higgs boson in the $\ell\ell bb$ final state in $pp$ collisions at $\sqrt{s}=13$ TeV with the ATLAS detector}",
    eprint = "1804.01126",
    archivePrefix = "arXiv",
    primaryClass = "hep-ex",
    reportNumber = "CERN-EP-2018-030",
    doi = "10.1016/j.physletb.2018.07.006",
    journal = "Phys. Lett. B",
    volume = "783",
    pages = "392--414",
    year = "2018"
}

@article{CMS:2019ogx,
    author = "Sirunyan, Albert M and others",
    collaboration = "CMS",
    title = "{Search for new neutral Higgs bosons through the H$\to$ ZA $\to \ell^{+}\ell^{-} \mathrm{b\bar{b}}$ process in pp collisions at $\sqrt{s} =$ 13 TeV}",
    eprint = "1911.03781",
    archivePrefix = "arXiv",
    primaryClass = "hep-ex",
    reportNumber = "CMS-HIG-18-012, CERN-EP-2019-254",
    doi = "10.1007/JHEP03(2020)055",
    journal = "JHEP",
    volume = "03",
    pages = "055",
    year = "2020"
}

@article{CMS:2019kca,
    author = "Sirunyan, Albert M and others",
    collaboration = "CMS",
    title = "{Search for a heavy pseudoscalar Higgs boson decaying into a 125 GeV Higgs boson and a Z boson in final states with two tau and two light leptons at $\sqrt{s}=$ 13 TeV}",
    eprint = "1910.11634",
    archivePrefix = "arXiv",
    primaryClass = "hep-ex",
    reportNumber = "CMS-HIG-18-023, CERN-EP-2019-231",
    doi = "10.1007/JHEP03(2020)065",
    journal = "JHEP",
    volume = "03",
    pages = "065",
    year = "2020"
}

@article{CMS:2019qcx,
    author = "Sirunyan, Albert M and others",
    collaboration = "CMS",
    title = "{Search for a heavy pseudoscalar boson decaying to a Z and a Higgs boson at $\sqrt{s} =$ 13 TeV}",
    eprint = "1903.00941",
    archivePrefix = "arXiv",
    primaryClass = "hep-ex",
    reportNumber = "CMS-HIG-18-005, CERN-EP-2018-343",
    doi = "10.1140/epjc/s10052-019-7058-z",
    journal = "Eur. Phys. J. C",
    volume = "79",
    number = "7",
    pages = "564",
    year = "2019"
}

@article{ATLAS:2020gxx,
    author = "Aad, Georges and others",
    collaboration = "ATLAS",
    title = "{Search for a heavy Higgs boson decaying into a Z boson and another heavy Higgs boson in the $\ell \ell bb$ and $\ell \ell WW$ final states in $pp$ collisions at $\sqrt{s}=13$ $\text {TeV}$ with the ATLAS detector}",
    eprint = "2011.05639",
    archivePrefix = "arXiv",
    primaryClass = "hep-ex",
    reportNumber = "CERN-EP-2020-191",
    doi = "10.1140/epjc/s10052-021-09117-5",
    journal = "Eur. Phys. J. C",
    volume = "81",
    number = "5",
    pages = "396",
    year = "2021"
}

@article{Carrington:1991hz,
    author = "Carrington, M. E.",
    title = "{The Effective potential at finite temperature in the Standard Model}",
    reportNumber = "TPI-MINN-91-48-T-REV, TPI-MINN-91-48-T",
    doi = "10.1103/PhysRevD.45.2933",
    journal = "Phys. Rev. D",
    volume = "45",
    pages = "2933--2944",
    year = "1992"
}

@article{Athron:2020sbe,
    author = "Athron, Peter and Bal{\'a}zs, Csaba and Fowlie, Andrew and Zhang, Yang",
    title = "{PhaseTracer: tracing cosmological phases and calculating transition properties}",
    eprint = "2003.02859",
    archivePrefix = "arXiv",
    primaryClass = "hep-ph",
    reportNumber = "CoEPP-MN-20-3",
    doi = "10.1140/epjc/s10052-020-8035-2",
    journal = "Eur. Phys. J. C",
    volume = "80",
    number = "6",
    pages = "567",
    year = "2020"
}

@article{Biekotter:2022kgf,
    author = {Biek{\"o}tter, Thomas and Heinemeyer, Sven and No, Jos{\'e} Miguel and Olea-Romacho, Mar{\'\i}a Olalla and Weiglein, Georg},
    title = "{The trap in the early Universe: impact on the interplay between gravitational waves and LHC physics in the 2HDM}",
    eprint = "2208.14466",
    archivePrefix = "arXiv",
    primaryClass = "hep-ph",
    reportNumber = "DESY-22-127, IFT--UAM/CSIC--22-015",
    doi = "10.1088/1475-7516/2023/03/031",
    journal = "JCAP",
    volume = "03",
    pages = "031",
    year = "2023"
}

@article{Dumont:2014wha,
    author = "Dumont, Beranger and Gunion, John F. and Jiang, Yun and Kraml, Sabine",
    title = "{Constraints on and future prospects for Two-Higgs-Doublet Models in light of the LHC Higgs signal}",
    eprint = "1405.3584",
    archivePrefix = "arXiv",
    primaryClass = "hep-ph",
    reportNumber = "LPSC14081, UCD-2014-03",
    doi = "10.1103/PhysRevD.90.035021",
    journal = "Phys. Rev. D",
    volume = "90",
    pages = "035021",
    year = "2014"
}

@article{Alwall:2014hca,
    author = "Alwall, J. and Frederix, R. and Frixione, S. and Hirschi, V. and Maltoni, F. and Mattelaer, O. and Shao, H. -S. and Stelzer, T. and Torrielli, P. and Zaro, M.",
    title = "{The automated computation of tree-level and next-to-leading order differential cross sections, and their matching to parton shower simulations}",
    eprint = "1405.0301",
    archivePrefix = "arXiv",
    primaryClass = "hep-ph",
    reportNumber = "CERN-PH-TH-2014-064, CP3-14-18, LPN14-066, MCNET-14-09, ZU-TH-14-14",
    doi = "10.1007/JHEP07(2014)079",
    journal = "JHEP",
    volume = "07",
    pages = "079",
    year = "2014"
}

@article{Sjostrand:2014zea,
    author = {Sj{\"o}strand, Torbj{\"o}rn and Ask, Stefan and Christiansen, Jesper R. and Corke, Richard and Desai, Nishita and Ilten, Philip and Mrenna, Stephen and Prestel, Stefan and Rasmussen, Christine O. and Skands, Peter Z.},
    title = "{An introduction to PYTHIA 8.2}",
    eprint = "1410.3012",
    archivePrefix = "arXiv",
    primaryClass = "hep-ph",
    reportNumber = "LU-TP-14-36, MCNET-14-22, CERN-PH-TH-2014-190, FERMILAB-PUB-14-316-CD, DESY-14-178, SLAC-PUB-16122",
    doi = "10.1016/j.cpc.2015.01.024",
    journal = "Comput. Phys. Commun.",
    volume = "191",
    pages = "159--177",
    year = "2015"
}

@article{deFavereau:2013fsa,
    author = "de Favereau, J. and Delaere, C. and Demin, P. and Giammanco, A. and Lema{\^\i}tre, V. and Mertens, A. and Selvaggi, M.",
    collaboration = "DELPHES 3",
    title = "{DELPHES 3, A modular framework for fast simulation of a generic collider experiment}",
    eprint = "1307.6346",
    archivePrefix = "arXiv",
    primaryClass = "hep-ex",
    doi = "10.1007/JHEP02(2014)057",
    journal = "JHEP",
    volume = "02",
    pages = "057",
    year = "2014"
}

@article{Czakon:2015qwa,
    author = {Czakon, Michael and Kr{\"a}mer, Michael and Worek, Malgorzata},
    editor = {Bl{\"u}mlein, Johannes and Jansen, Karl and Kr{\"a}mer, Michael and K{\"u}hn, Johann H.},
    title = "{Automated NLO/NLL Monte Carlo programs for the LHC}",
    eprint = "1502.01521",
    archivePrefix = "arXiv",
    primaryClass = "hep-ph",
    reportNumber = "TTK-15-03",
    doi = "10.1016/j.nuclphysbps.2015.03.009",
    journal = "Nucl. Part. Phys. Proc.",
    volume = "261-262",
    pages = "93--114",
    year = "2015"
}

@article{Sonnenschein:2006ud,
    author = "Sonnenschein, Lars",
    title = "{Analytical solution of ttbar dilepton equations}",
    eprint = "hep-ph/0603011",
    archivePrefix = "arXiv",
    doi = "10.1103/PhysRevD.78.079902",
    journal = "Phys. Rev. D",
    volume = "73",
    pages = "054015",
    year = "2006",
    note = "[Erratum: Phys.Rev.D 78, 079902 (2008)]"
}

@article{Arco:2025ydq,
    author = {Arco, Francisco and Biek{\"o}tter, Thomas and Stylianou, Panagiotis and Weiglein, Georg},
    title = "{Top-quark spin correlations as a tool to distinguish pseudoscalar A {\textrightarrow} ZH and scalar H {\textrightarrow} ZA signatures in $ Zt\overline{t} $ final states at the LHC}",
    eprint = "2502.03443",
    archivePrefix = "arXiv",
    primaryClass = "hep-ph",
    reportNumber = "IFT--UAM/CSIC-24-187, DESY-25-023",
    doi = "10.1007/JHEP06(2025)170",
    journal = "JHEP",
    volume = "06",
    pages = "170",
    year = "2025"
}

@article{TMVA:2007ngy,
    author = "Hocker, Andreas and others",
    collaboration = "TMVA",
    title = "{TMVA - Toolkit for Multivariate Data Analysis}",
    eprint = "physics/0703039",
    archivePrefix = "arXiv",
    reportNumber = "CERN-OPEN-2007-007",
    month = "3",
    year = "2007"
}

@article{PhysRevD.110.030001,
  title = {Review of Particle Physics},
  author = {Navas, S et al.},
  collaboration = {Particle Data Group Collaboration},
  journal = {Phys. Rev. D},
  volume = {110},
  issue = {3},
  pages = {030001},
  numpages = {5},
  year = {2024},
  month = {Aug},
  publisher = {American Physical Society},
  doi = {10.1103/PhysRevD.110.030001},
  url = {https://link.aps.org/doi/10.1103/PhysRevD.110.030001}
}

@article{Brun:1997pa,
    author = "Brun, Rene and Rademakers, Fons",
    editor = "Werlen, M. and Perret-Gallix, D.",
    title = "{ROOT {\textemdash} An object oriented data analysis framework}",
    doi = "10.1016/s0168-9002(97)00048-x",
    journal = "Nucl. Instrum. Meth. A",
    volume = "389",
    number = "1-2",
    pages = "81--86",
    year = "1997"
}

@article{Curtin:2016urg,
    author = "Curtin, David and Meade, Patrick and Ramani, Harikrishnan",
    title = "{Thermal Resummation and Phase Transitions}",
    eprint = "1612.00466",
    archivePrefix = "arXiv",
    primaryClass = "hep-ph",
    reportNumber = "YITP-2016-48",
    doi = "10.1140/epjc/s10052-018-6268-0",
    journal = "Eur. Phys. J. C",
    volume = "78",
    number = "9",
    pages = "787",
    year = "2018"
}

@article{Bernreuther:2015yna,
    author = "Bernreuther, Werner and Heisler, Dennis and Si, Zong-Guo",
    title = "{A set of top quark spin correlation and polarization observables for the LHC: Standard Model predictions and new physics contributions}",
    eprint = "1508.05271",
    archivePrefix = "arXiv",
    primaryClass = "hep-ph",
    reportNumber = "TTK-15-16",
    doi = "10.1007/JHEP12(2015)026",
    journal = "JHEP",
    volume = "12",
    pages = "026",
    year = "2015"
}

@article{Bechtle:2020pkv,
    author = "Bechtle, Philip and Dercks, Daniel and Heinemeyer, Sven and Klingl, Tobias and Stefaniak, Tim and Weiglein, Georg and Wittbrodt, Jonas",
    title = "{HiggsBounds-5: Testing Higgs Sectors in the LHC 13 TeV Era}",
    eprint = "2006.06007",
    archivePrefix = "arXiv",
    primaryClass = "hep-ph",
    reportNumber = "BONN-TH-2020-03, DESY 20-093, DESY-20-093, IFT-UAM/CSIC-20-072, LU 20-27",
    doi = "10.1140/epjc/s10052-020-08557-9",
    journal = "Eur. Phys. J. C",
    volume = "80",
    number = "12",
    pages = "1211",
    year = "2020"
}

@article{Bechtle:2020uwn,
    author = "Bechtle, Philip and Heinemeyer, Sven and Klingl, Tobias and Stefaniak, Tim and Weiglein, Georg and Wittbrodt, Jonas",
    title = "{HiggsSignals-2: Probing new physics with precision Higgs measurements in the LHC 13 TeV era}",
    eprint = "2012.09197",
    archivePrefix = "arXiv",
    primaryClass = "hep-ph",
    reportNumber = "BONN-TH-2020-09, DESY-20-228, DESY 20-228, IFT-UAM/CSIC-20-081, LU TP 20-53",
    doi = "10.1140/epjc/s10052-021-08942-y",
    journal = "Eur. Phys. J. C",
    volume = "81",
    number = "2",
    pages = "145",
    year = "2021"
}

\end{document}